\documentclass{elsart5p}

\usepackage{graphicx}

\usepackage{amssymb}
\usepackage{amsmath}
\usepackage{amsfonts}
\usepackage{longtable}
\usepackage{url}
\usepackage{rotating}
\usepackage{lscape}
\usepackage[margin=0pt,font=small,labelfont=rm,labelsep=period,format=hang,position=top]{caption}
\usepackage{hyperref} 

\journal{Physica A: Statistical Mechanics and its Applications}

\newcommand{\N}{\mathbb{N}}
\newcommand{\Z}{\mathbb{Z}}

\newcommand{\R}{\mathbb{R}}
\newcommand{\E}{\mathbb{E}}

\newcommand{\Hss}{\textit{H-ss}}

\newcommand{\fbm}{\textit{fBm}}
\newcommand{\fgn}{\textit{fGn}}

\newcommand{\ckd}{\qquad\square}

\begin{document}

\begin{frontmatter}

\title{Blocks adjustment --- reduction of bias and variance of detrended fluctuation analysis using Monte Carlo simulation}
\author[label1]{Sebastian Michalski}
\address[label1]{Institute of Econometrics, Warsaw School of Economics, Al. Niepodleg\l{}o\'sci 162, 02-554 Warsaw, Poland}
\ead{smicha@sgh.waw.pl}

\begin{abstract}
The length of minimal and maximal blocks equally distant on log-log scale versus fluctuation function considerably influences bias and variance of DFA. Through a number of extensive Monte Carlo simulations and different fractional Brownian motion/fractional Gaussian noise generators, we found the pair of minimal and maximal blocks that minimizes the sum of mean-squared error of estimated Hurst exponents for the series of length $N=2^p, p=7,\ldots,15$. Sensitivity of DFA to sort-range correlations was examined using ARFIMA($p,d,q$) generator. Due to the bias of the estimator for anti-persistent processes, we narrowed down the range of Hurst exponent to $1/2\leq H< 1.$
\end{abstract}

\begin{keyword}
Detrended Fluctuation Analysis \sep Scaled Windowed Variance \sep fractional Brownian motion \sep Hurst exponent \sep ARFIMA
\end{keyword}
\end{frontmatter}

\section{Introduction}

As of September, 2004, the two original papers \cite{peng94,peng95} on Detrended Fluctuation Analysis (DFA) had been cited by 470 research articles, and the number is still growing.\footnote{\url{www.physionet.org/physiotools/dfa/citations.shtml}} But still there is a need for improving the methodology, e.g.,~testing the limitations of DFA for various types of nonstationarities \cite{chen02}, investigating its performance for signals with different type of correlations, with random spikes and missing segments, comparing it with other methods \cite{xu05,mielwojd2007}, testing the effects of periodic (and quasi-periodic) trends in the estimation process \cite{hu01,phA:354:182}, studying the limitations of DFA for anti-persistent signals and the strategies to overcome them \cite{chen05}, its application to the wide class of multifractal series \cite{oswiecimka06} or latest works on an impact of coarse-graining \cite{phA:363:226}. The method is also known as a Scaled Windowed Variance -- Linear Detrended \cite{mandelbrot85}, Roughness Around the Root Mean Square Line \cite{moreira94} and Residuals of Regression \cite{taqqu96}. 

Thorough examination of DFA as a type of linear regression detrended Scaled Windowed Variance method was carried out by \cite{phA:241:606} and later by \cite{phA:265:85}.\footnote{First type of Scaled Windowed Variance method known as the bridge method \cite{phA:241:606} was proposed by \cite{mandelbrot85}.} Authors indicated sensitivity of DFA to exclusion of blocks of different size and after simulation stated that
\begin{quote}
,,\ldots excluding windows of large sizes reduces variance but results in significant bias, while excluding windows of small sizes reduces bias in estimates nearly to zero for all signal lengths and all values of true $H$ but the variance increases dramatically.''
\end{quote}
In this study we will try to find the best block cuts by conducting exhaustive experiments with 10,000 replications using different types of fractional Brownian motion or fractional Gaussian noise generators. Final results will be verified under the presence of short-range dependence using ARIFMA$(p,d,q)$ process.

\section{Some preliminaries}

Self-similar real-valued stochastic process $X=\{X(t)\}_{t\in\R}$ with Hurst exponent $H>0$ (\Hss) is defined as satisfying $\{X(at)\}_{t\in\R}\stackrel{d}{=}\{a^HX(t)\}_{t\in\R}$, for $a>0$. Hyperbolically-decaying autocorrelation function $\gamma(k)$ of a stationary stochastic process $\{X_t\}_{t=0}^{\infty}$ is nonsummable (i.e. $\sum_k\gamma(k)=\infty$) and defines \textit{asymptotically} self-similar process if
\begin{equation}
	\gamma(k)\propto k^{2H-2}L(k), \quad k\to\infty, \quad 1/2<H<1,
	\label{eq:asympss}
\end{equation}
where $L(k)$ is a slowly-varying function, i.e. $\lim_{t\to\infty}L(tk)/L(k)=1$, and defines \textit{exactly} self-similar process if 
\begin{equation}
	\gamma(k)=1/2[(k+1)^{2H}-2k^{2H}+(k-1)^{2H}].
	\label{eq:exactlyss}
\end{equation}
If $\gamma(k)$ is diverging, one says that $\{X_t\}_{t\in\Z}$ exhibits \textit{long-range dependence} (LRD), \textit{strong dependence}, has \textit{long memory} or is a $1/f$ \textit{noise} \cite{doukhan03}. LRD corresponds to the blow-up of the spectral density $S(f)$ at the origin
\begin{equation}
	S(f)\propto cf^{1-2H}, \quad  f\to 0, \quad 0<H<1,
	\label{eq:powerlrd}
\end{equation}
where $\{S(f)\}_{f\in[-\pi,\pi]}$
\begin{equation}
	S(f)=\frac{1}{2\pi}\sum_{k=-\infty}^{\infty}e^{-ifk}\gamma(k).
	\label{eq:powerlrd2}
\end{equation}
Partitioning self-similar process $\{X_i\}_{i\in\Z}$ into non-overlapping blocks of $m$ sequential elements and computing average of these $m$ elements
\begin{equation}
	X_t^{(m)}=\frac{1}{m}\sum_{i=(t-1)m+1}^{tm}X_i
	\label{eq:partition}
\end{equation}
does not change the autocorrelation function \cite{paxson97} (contrary to ''typical'' stochastic processes --- $m$ increases and autocorrelation of $\{X_t^{(m)}\}$ decreases).

The most widely-studied self-similar processes are fractional Gaussian noise (fGn) \cite{mandelbrotvan68} and autoregressive fractional integrated moving average processes (ARFIMA) \cite{granger80,hosking81}. In this study we used $ARFIMA(0,d,0)$, to generate \Hss. The general $ARFIMA(p,d,q)$ process is defined as
\begin{equation}
	\Phi(L)(1-L)^dX_t=\Theta(L)\epsilon_t,
	\label{eq:arfimapdq}
\end{equation}
where $L$ is the lag operator, $\epsilon_t$ is white noise process and $d\in\R$ is the fractional differencing parameter such that $|d|<1/2$. The process is covariance stationary if $-1/2<d<1/2$ and invertible for $d>-1/2$. For $p,q=0$ we have $(1-L)^dX_t=\epsilon_t$, and its Wold representation is given by
\begin{equation}
	X_t=\sum_{j=0}^{\infty}\pi_jL^j,
	\label{eq:woldarfima}
\end{equation}
where $\pi_0=1$ and 
\begin{equation}
	\pi_j=\prod_{k=1}^{j}\frac{k-1-d}{k}=\frac{\Gamma(j-d)}{\Gamma(j+1)\Gamma(-d)}, \quad j\in\N.
	\label{eq:woldarfima2}
\end{equation}
Covariance function $\gamma(k)=\E X_0X_k$ of $ARFIMA(0,d,0)$, $-1/2<d<1/2$ is given by \cite{doukhan03}
\begin{equation}
	\gamma(k)=\sigma^2\frac{(-1)^k\Gamma(1-2d)}{\Gamma(k-d+1)\Gamma(1-k-d)}
	=\sigma^2\frac{\Gamma(k+d)\Gamma(1-2d)}{\Gamma(k-d+1)\Gamma(d)\Gamma(1-d)}
	\propto c|k|^{2d-1} \text{ as $k\to\infty$}. 
	\label{eq:arfimacov}
\end{equation}
Hence, $ARFIMA(0,d,0)$ has long memory if and only if $0<d<1/2$. If we compare Eq. \eqref{eq:asympss} and \eqref{eq:arfimacov} we see that $H=d+1/2$. If $\{X_t\}_{t\in\Z}$ is a Gaussian $ARFIMA(0,d,0)$, $0<d<1/2$, then as $n\to\infty$ $n^{-H}\sum_{t=1}^{\lfloor ns \rfloor}X_t\to B_H(s)$, where $\{B_H(s)\}_{s\in\R}$ is a fractional Brownian motion.

\section{Detrended Fluctuation Analysis}

Let $X_i$ be the stationary series of compact support, where support is defined as a set of indices $i$ with nonzero values $X_i$. The series is compact if $X_i=0$ for its small fraction only, interpreted then as having no value at this $i$.

First the series is divided into $\lfloor N/m \rfloor$ non-overlapping\footnote{Overlapping blocks introduce correlations between estimates and should be abandoned.} logarithmically spaced blocks (windows, segments) of size $m$. Since $N$ is often not a multiple of time scale $m$, a short part at the end of the series may remain. In order not to disregard this part of the series, the same procedure is repeated starting from the opposite end, so $2N/m$ segments are obtained altogether. In this study we set $N$ (and $m$) to a power of 2, so that $N/m$ is directly integer and a final number of blocks.

The advantage of DFA is that it is applied directly to nonstationary series. Hence, as we assumed $X_i$ is stationary, the series must be integrated before analysis, calculating within each block partial sums $Y(t)\equiv\sum_{i=1}^{t}X_i$. During integration the sample mean $\bar X=N^{-1}\sum_{i=1}^NX_i$ can be subtracted (adjusted partial sums) $Y(t)\equiv\sum_{i=1}^{t}(X_i-\bar X)$ but not compulsory because it will be eliminated by the later detrending. Within each $k=1,2,\ldots,N/m$ block a least square line, $a_k+b_kt$, is fitted to the partial sums, and the sample variance of residuals is computed
\begin{equation}
F^2(k,m)\equiv\frac{1}{m-1}\sum_{t=1}^{m}\bigg(Y((k-1)m+t)-a_k-b_kt\bigg)^2, \quad k=1,2,\ldots,N/m.
\label{eq:1}
\end{equation}
One can switch the fitting trend to quadratic, cubic, or higher order polynomials (called DFA2, DFA3, \ldots, DFA$r$) \cite{peng94} and by comparing the results for different $r$ estimate the type of the polynomial trend in the time series \cite{phA:295:441}. Averaging Eq. \eqref{eq:1} over all blocks gives the $q$th order fluctuation function
\begin{equation}
F_q(m)\equiv\left\{\frac{1}{N/m}\sum_{k=1}^{N/m}\left(F^2(k,m)\right)^{q/2}\right\}^{1/q}
\label{eq:2}
\end{equation}
which is by construction defined only for $m \ge r+2$. Eq. \eqref{eq:2} refers to multifractal detrended fluctuation analysis (MF-DFA) \cite{phA:316:87}, thoroughly analyzed by Ref. \cite{oswiecimka06}. In this study we focus only on the standard DFA, i.e. $q=2$
\begin{equation}
F(m)\equiv\left\{\frac{1}{N/m}\sum_{k=1}^{N/m}\left(F^2(k,m)\right)\right\}^{1/2}
\label{eq:3}
\end{equation}
The variance of residuals is proportional to $m^{2H}$, where $H$ is Hurst exponent \cite{taqqu98}, \cite{taqqu96}. Hence the fluctuation function described in Eq. \eqref{eq:3} is proportional to $m^H$
\begin{equation}
F(m)\propto m^H.
\label{eq:4}
\end{equation}
The scaling behavior of the fluctuation function is analyzed on log-log plots $F(m)$ versus $m$ -- the slope of regression $\log(F(m))=c+H\log m$ is the Hurst exponent.
As will be shown further DFA can reliably determine persistent signals ($1/2<H<1$), and it becomes inaccurate for strongly anti-persistent processes ($0<H<1/2$) when $H$ is close to zero. In such cases, a modified DFA technique has to be used.  The easiest way is to apply --- instead of single summation --- double summation
\begin{equation}
\tilde{Y}(j) \equiv \sum_{t=1}^j \left(Y(t)-\bar Y \right).
 \label{eq:5}
\end{equation}
It leads to so called generalized fluctuation functions $\tilde{F}(m)$ described by a scaling law with larger than in Eq. \eqref{eq:4} exponents
\begin{equation}
\tilde{F}(m)\propto m^{\tilde{H}} = m^{H+1}.
	\label{eq:6}
\end{equation}
Comparing Eq. \eqref{eq:4} and \eqref{eq:6} we see that $\tilde{F}(m)/m=F(m)$. Nevertheless double summation leads to quadratic trends in $\tilde{F}(m)$. Hence if the average values were not removed in Eq. \eqref{eq:5}, at least the second order DFA should be applied to eliminate these artificial trends. Due to these inconveniences (and greater occurrence of persistent signals) we narrow down our analysis to persistent processes.

The sum of the numbers $X_i$ within each block $k$ of size $m$ is known as the box probability $p_m(k)$ in the standard multifractal formalism for normalized series
\begin{equation}
p_m(k)\equiv\sum_{i=(k-1)m+1}^{km}X_i,\quad k=1,2,\ldots,N/m,
	\label{eq:7}
\end{equation}
and defines partition function with scaling exponent $\tau(q)=qh(q)-1$ (here for $q=2$)
\begin{equation}
Z_q(m)\equiv\sum_{k=1}^{N/m}|p_m(k)|^q~\propto m^{qh(q)-1}=m^{2H-1}.
	\label{eq:8}
\end{equation}
We can relate the scaling exponent from Eq. \eqref{eq:8} to H\"older exponent $\alpha$ and singularity spectrum $f(\alpha)$

\begin{equation}
f(\alpha)=2(\alpha-H)+1.
	\label{eq:9}
\end{equation}

\section{Monte Carlo simulation}\label{sec:sim}

\noindent
In order to carry out the experiment we have chosen the following fractional Brownian motion generators:
\begin{itemize}
	\item Davies and Harte (known also as Wood and Chan, circulant matrix embedding method, exact) \cite{davies87}, \cite{doukhan03},
	\item Hosking method (recursive, exact, known also as Levinson method for Toeplitz matrices) \cite{hosking84}, \cite{coeurjolly00},
	\item Choleski decomposition of the covariance matrix (exact) \cite{coeurjolly00}, \cite{doukhan03},
\end{itemize}
and fractional Gaussian noise generators:
\begin{itemize}
	\item Paxson method (approximate) \cite{paxson97},
	\item Beran method \cite{beran94},
	\item Durbin--Levinson (using Yule--Walker-type equations \cite{brockwell91}), \cite{taqqu96}, \cite{adler},
	\item ARFIMA$(0,d,0)$ (in the frequency domain using fast Fourier transform based on S-PLUS code written originally by \cite{beran94}).
\end{itemize}
All fractional Gaussian noise series were cumulated before $H$ estimation. Sensitivity of final results on short-range dependence was examined for ARFIMA$(p,d,q)$ series: $(0,d,1)$ with $\theta=0.5$, $(1,d,0)$ with $\phi=0.5$, $(1,d,1)$ with $\phi=0.3,\theta=0.7$, $(1,d,1)$ with $\phi=-0.3,\theta=-0.7$ and $(1,d,1)$ with $\phi=0.7,\theta=0.3$. 

\begin{figure}[b!]
	\centering
		\includegraphics[width=.54\textwidth]{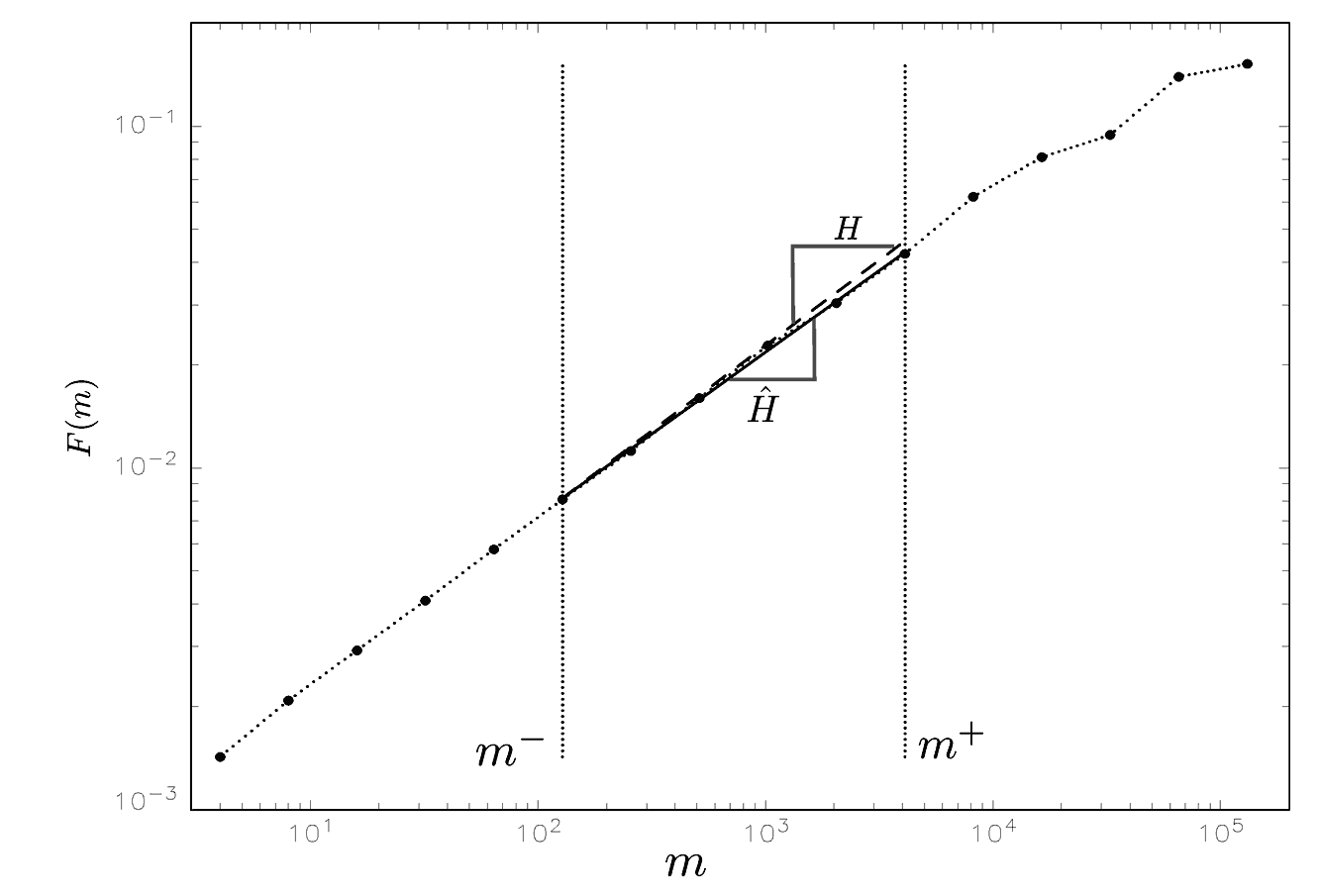}
	\caption{An example of the fluctuation function $F(m)$ and minimal and maximal blocks $m^-$, $m^+$ (left and right dotted vertical lines). The slope of the solid line represents $\hat{H}$, the slope of the dashed line -- nominal $H$.}
	\label{fig:m1m2}
\end{figure}

Because $H$ is not known \textit{a priori}, suggestions for block adjustment is based only on the length of the series. The optimization criterion has to be chosen to minimize both bias and variance for known $N$. Hence, a sum of mean-squared error (MSE) was chosen as the criterion of method reliability. Minimizing MSE accounts for square of bias and variance minimization \cite{maddala06}
\begin{equation}
	\textup{MSE}(\hat H)\equiv \E(\hat H-H)^2=(\E\hat H-H)^2+\E(\hat H-\E \hat H)^2=bias_{\hat H}^2+variance_{\hat H}
	\label{eq:mse}
\end{equation}
\emph{Proof}
\begin{align*}\label{eq:mseproof}
	\textup{MSE}(\hat H)\equiv \E(\hat H-H)^2&=\E(\hat H^2+H^2-2H\hat H)\\
																	&=\E(\hat H^2)+H^2-2H\E \hat H\\
																	&=\E(\hat H^2)+H^2-2H\E \hat H+(\E \hat H)^2+(\E \hat H)^2-2(\E \hat H)^2\\
																	&=(\E \hat H)^2+H^2-2H\E \hat H+\E(\hat H^2)+(\E \hat H)^2-2(\E \hat H)^2\\
																	&=(\E \hat H)^2+H^2-2H\E \hat H+\E(\hat H^2)+\E(\E \hat H)^2-2\E \hat H \E \hat H\\
																	&=(\E \hat H)^2+H^2-2H\E \hat H+\E(\hat H^2+(\E \hat H)^2-2\hat H \E \hat H)\\
																	&=(\E\hat H-H)^2+\E(\hat H-\E \hat H)^2 \ckd																	
\end{align*}
Let us consider MSE$(\hat H)$ as a function of a pair of minimal and maximal blocks $(m^-,m^+)$ chosen in the estimation process (Figure~\ref{fig:m1m2}). Assuming, that DFA is constructed on at least $c^*$ different equally distant on log-log scale blocks, we can define the set of all possible combinations of $(m^-,m^+)$ as
\begin{equation}
	\mathcal{C}\equiv\{(m^-,m^+)=(2^l,2^u)\colon u-l+1 \ge c^* \wedge l=l_1,\ldots,\log_2N \wedge u=u_1,\ldots,\log_2N \}.
	\label{eq:comb}
\end{equation}

\clearpage

\small
\begin{table}[t!]
\caption{Number of elements of $\mathcal{C}$ for series of length $N=2^7,\ldots,2^{15}$ for different order of fitting polynomial trend $r$. For linear ($r=1$) and quadratic ($r=2$) polynomial trend $l_1=2$. For cubic ($r=3$) and 4th order polynomial fitting trend ($r=4$) $l_1=3$. $c^*=4$.}
\label{tab:noofc}
\begin{tabular}{rrrrr}
\hline
	$p$ & $N=2^p$&\multicolumn{3}{c}{$\#\mathcal{C}$}\\
	\hline
	   &        &\multicolumn{3}{c}{fitting trend}\\
	\cline{3-5}
	   &        &linear or quadratic&&cubic or 4th order\\
	\cline{3-3}\cline{5-5}
	 7 &    128 &  6 &&  3 \\
	 8 &    256 & 10 &&  6 \\
	 9 &    512 & 15 && 10 \\
	10 &  1~024 & 21 && 15 \\
	11 &  2~048 & 28 && 21 \\
	12 &  4~096 & 36 && 28 \\
	13 &  8~192 & 45 && 36 \\
	14 & 16~384 & 55 && 45 \\
	15 & 32~768 & 66 && 55 \\
\hline
\end{tabular}
\end{table}
\normalsize

\noindent
The number of all elements of $\mathcal{C}$ (presented in Table \ref{tab:noofc}) is then given by
\begin{equation}
	\#\mathcal{C}=a(a+1)/2,\text{ where } a=\log_2N+2-l_1-c^*.
	\label{eq:noofcomb}
\end{equation}
If the fitting trend is linear or quadratic (see Eq.~\eqref{eq:1}) we set the shortest block to $2^{l_1}=4$. For cubic and 4th order polynomial trend we set it to $2^{l_1}=8$. The minimal number of different blocks is set to $c^*=4$.
\noindent
Let us also introduce the following function
\begin{equation}
	\vartheta(m^-,m^+)\equiv\sum_{H\in\mathcal{H}} \text{MSE}(\hat H)(m^-,m^+),
	\label{eq:summse}
\end{equation}
which idea is to describe the behavior of MSE$(\hat H)$ for an \emph{a priori} unknown Hurst parameter. Due to the strong bias of DFA for anti-persistent processes \cite{peng95} we have considered in our simulation persistent processes only. Of course MSE$(\hat H)$ is sensitive to length of series $N$ and $H$, but to make the notation in Eq.~\eqref{eq:summse} clear we omitted them. Hence, all values of $\vartheta$ that will appear in our study are computed on $\mathcal{H}=\{0.5,0.6,0.7,0.8,0.9\}$. Our goal is to find --- through the number of computer simulations --- the pair of $(\tilde{m}^-,\tilde{m}^+)$, which will minimize $\vartheta$ for long memory time series of length $N$
\begin{equation}
	(\tilde{m}^-,\tilde{m}^+)=\arg\min\vartheta(m^-,m^+).
	\label{eq:argmin}
\end{equation}
Before this step we tried approximate the number of the series iterations (replications), under which $(\tilde{m}^-,\tilde{m}^+)$ would not change (within a small error). Hence, for a fixed number of replications of the $\fbm$ or $\fgn$ series of length~$N$, we calculated $(\tilde{m}^-,\tilde{m}^+)$ and then looped this step 50 times, each time recording $(\tilde{m}^-,\tilde{m}^+)_{ex}$, where $ex$ indicates subsequent experiments. Then from these 50~optimal pairs of minimal and maximal blocks we calculated the mode and the frequency at which it occurred (in these 50 experiments)
\begin{equation}
	P(\tilde{m}^-,\tilde{m}^+)=\max_{(\tilde{m}^-,\tilde{m}^+)\in\mathcal{C}} P\left((\tilde{m}^-,\tilde{m}^+)_{ex}\right), \quad ex=1,\ldots,50,
	\label{eq:pmode}
\end{equation}
and next we increased the number of replications. In Table~\ref{tab:mode} there is an example of the way $P(\tilde{m}^-,\tilde{m}^+)$ is determined for the series of length $N=128$ iterated certain times.
\small
\begin{table}[b!]
\caption{Example of the construction of $P(\tilde{m}^-,\tilde{m}^+)$ for $N=128$. Before the simulation is carried out we list $\mathcal{C}$, which has six possible block combinations $(m^-,m^+)$. After each of 50 experiments, $(m^-,m^+)$ are sorted by $\vartheta$ in ascending order and $P(\tilde{m}^-,\tilde{m}^+)$ for the best (bold font) combination $(\tilde{m}^-,\tilde{m}^+)$ is computed.}
\label{tab:mode}
\begin{tabular}{rrrrrrrrrrrrrrrccccrrrrrrrr}
\hline
\multicolumn{4}{c}{$\mathcal{C}$}&&&&\multicolumn{14}{c}{after experiment}&&&&&&\\
\cline{8-21}
\multicolumn{4}{c}{}&&&&\multicolumn{2}{c}{1st}&&&&\multicolumn{2}{c}{2nd}&&&$\cdots$&&&\multicolumn{2}{c}{50th}&&&&\multicolumn{2}{c}{mode}&$P(\tilde{m}^-,\tilde{m}^+)$\\
&$(m^-$,&$m^+)$&&&&&$(m^-$,&$m^+)$&&&&$(m^-$,&$m^+)$&&&&&&$(m^-$,&$m^+)$&&&&$(\tilde{m}^-$,&$\tilde{m}^+)$&\\
\cline{1-4}\cline{8-9}\cline{13-14}\cline{20-21}\cline{25-26}
&(4,&128)&&&&&({\bf4},&{\bf32})&&&&({\bf4},&{\bf64})&&&$\cdots$&&&({\bf4},&{\bf32})&&&&(4,&32)&0.48\\
&(4,&64)&&&&&(4,&64)&&&&(4,&32)&&&$\cdots$&&&(8,&128)&&&&&&\\
&(4,&32)&&&&&(4,&128)&&&&(8,&64)&&&$\cdots$&&&(4,&128)&&&&&&\\
&(8,&128)&&&&&(8,&128)&&&&(4,&128)&&&$\cdots$&&&(4,&64)&&&&&&\\
&(8,&64)&&&&&(16,&128)&&&&(8,&128)&&&$\cdots$&&&(16,&128)&&&&&&\\
&(16,&128)&&&&&(8,&64)&&&&(16,&128)&&&$\cdots$&&&(8,&64)&&&&&&\\
\hline
\end{tabular}
\end{table}
\normalsize
\clearpage

\begin{figure}[t!]
	\centering
		\includegraphics[width=.5\textwidth]{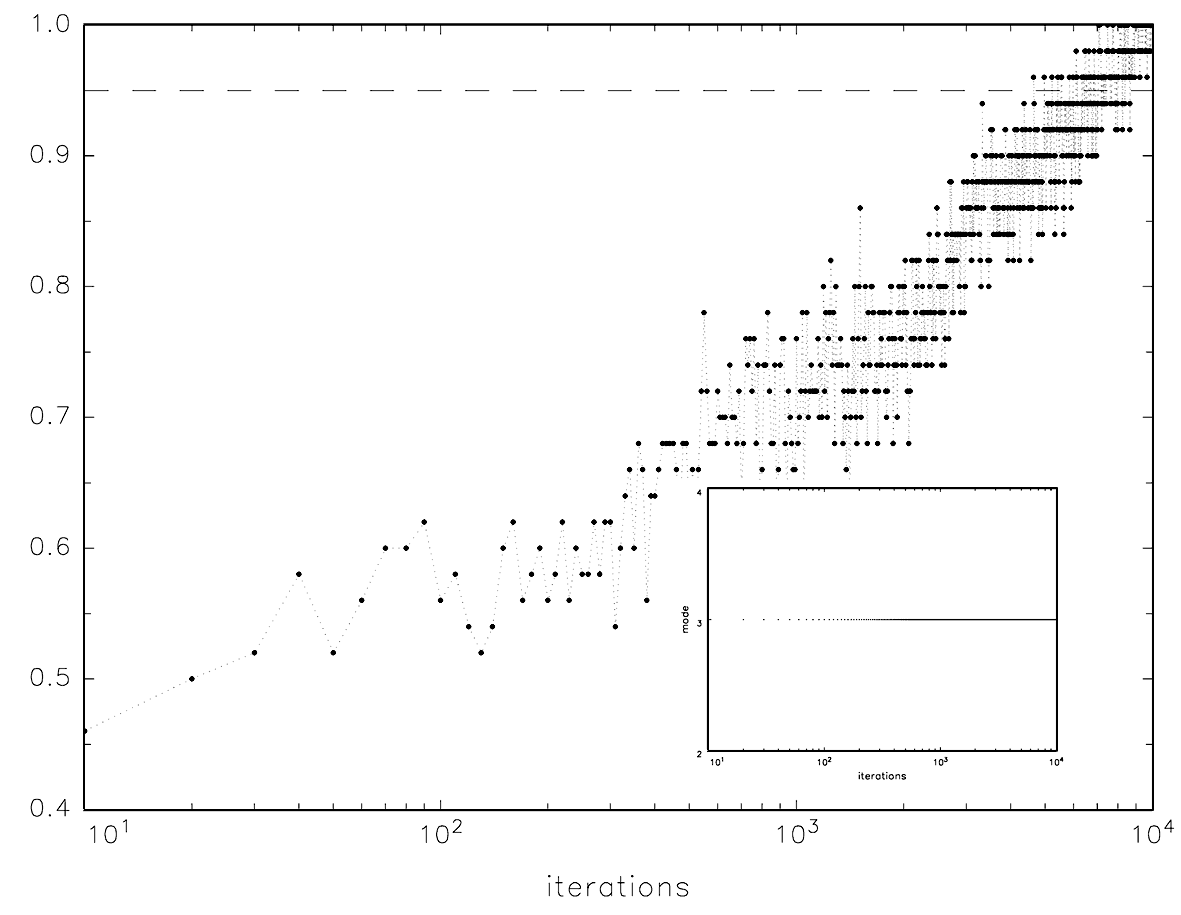}
	\caption{$P(\tilde{m}^-,\tilde{m}^+)$ for $N=128$ and 50 experiments against increasing iterations (by 10). While the first point is based on 250 \emph{fBm} simulated series (10 iterations $\times$ 50 experiments $\times$ 5 Hurst exponents), the last one --- on 2,500,000. Inner panel shows stability of mode (3~is $(\tilde{m}^-,\tilde{m}^+)=(4,32)$). \emph{fBm} generator: Davies and Harte.}
	\label{fig:mode}
\end{figure}

Figure~\ref{fig:mode} depicts the behavior of $P(\tilde{m}^-,\tilde{m}^+)$ against the number of replications from 10 to 10,000 (increased by 10) and looped 50~times. Note that the last point on Figure~\ref{fig:mode} represents 2,500,000 generations of \emph{fBm} of length $N=128$ (10,000~iterations $\times$ 50~experiments $\times$ 5~Hurst parameters). We see that for 100 iterations $P(\tilde{m}^-,\tilde{m}^+)$ is about 0.60, which means that only 60\% of 50 carried experiments gave the same $(\tilde{m}^-,\tilde{m}^+)$. For 1,000 iterations $P(\tilde{m}^-,\tilde{m}^+)$ reaches about 0.75 and for 10,000 is above~0.95. At this number of iterations there exists a small (less than 5\%) error/risk that obtained $(\tilde{m}^-,\tilde{m}^+)$  will change if the experiment is repeated. Hence, we will carry out our computations of $\vartheta(\tilde{m}^-,\tilde{m}^+)$ simulating 10,000 replications of LRD series for each of $H\in\{0.5,0.6,0.7,0.8,0.9\}$. For the purpose of a graphical presentation of the behavior of DFA $\mathcal{H}=\{0.1,0.2,\ldots,0.9\}$ has been chosen. Additionaly we will list up to three the best block combinations for different $N$ and $H$.

\section{Results and discussion}\label{sec:res}

Table~\ref{tab:fbmfgn_all} presents the best three minimal and maximal block combinations $(\tilde{m}^-,\tilde{m}^+)$ (notation $(\tilde{m}^-,\tilde{m}^+)$ is reserved for \#1) for 10,000 \emph{fBm}/\emph{fGn} independent paths generated with the use of different generators listed in section \ref{sec:sim}. Despite the differences in used generators, in five out of six the pair $(4,32)$ minimizes $\vartheta(m^-,m^+)$, the second and third best pairs are $(4,64)$ and $(4,128)$ respectively. The only exception is the result obtained using Hosking generator, but the pairs $(4,32)$, $(4,64)$, $(4,128)$ are still preferable. We also see that the $(\tilde{m}^-,\tilde{m}^+)$ do not change with the increase of the number of observations $N$.

\begin{figure}[t!]
	\centering
		\includegraphics[width=.7\textwidth]{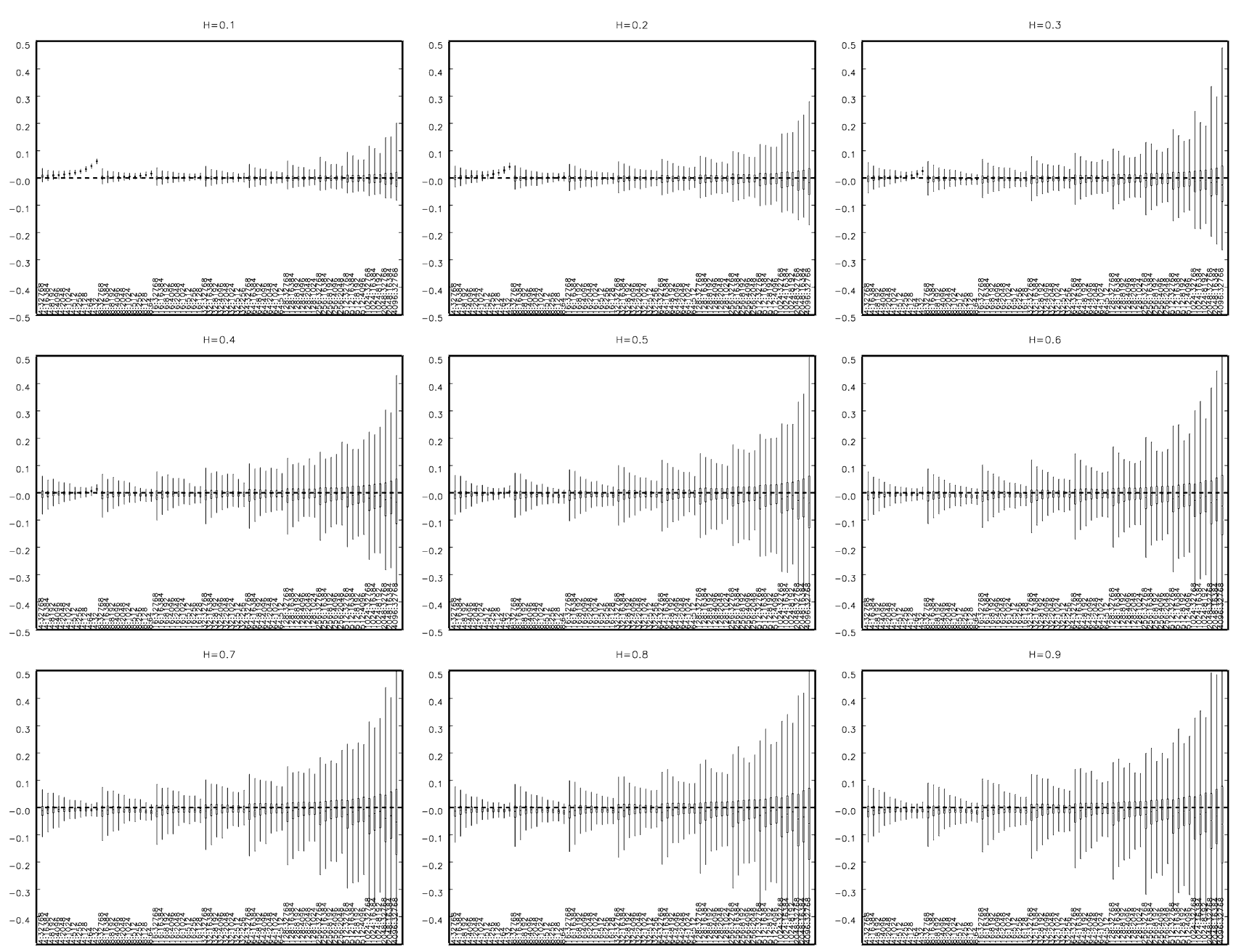}
	\caption{Boxplots for 10,000 \emph{fBm} paths of length $N=32768$ generated using Davies and Harte exact method. On X axis --- different block combinations, starting from 4 obs. in the shortest block and $N$ in the longest one. Longest blocks are cut first till at least four blocks are left. On Y axis --- deviation of the nominal value.}
	\label{fig:2p15circ}
\end{figure}

\begin{figure}[t!]
	\centering
		\includegraphics[width=.6\textwidth]{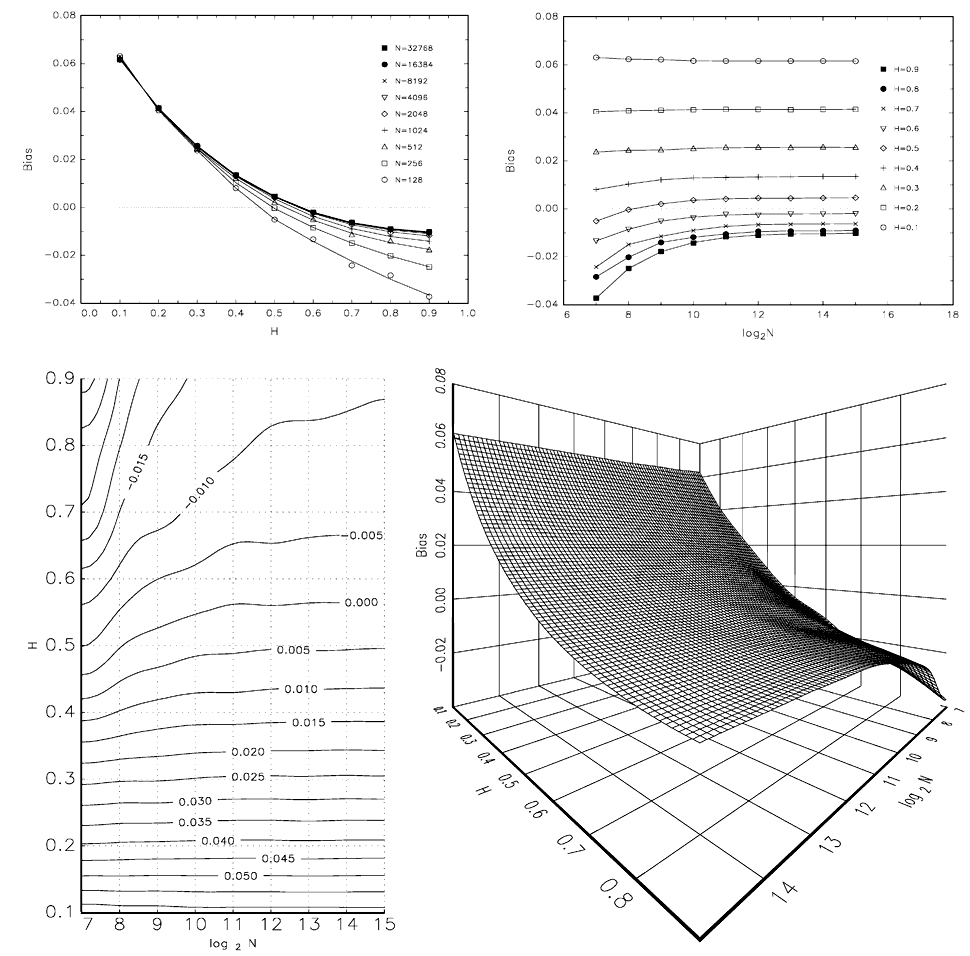}
	\caption{Upper left: scatter plot of bias against $H$ with fitted lines for different length of the series $N=2^7,\ldots,2^{15}$. Upper right: scatter plot of bias against $\log_2N$ with fitted lines for different Hurst parameters, $H=0.1,\ldots,0.9$. Lower right: cubic spline fit of bias with $H\times\log_2N$. Lower left: contour of cubic spline fitted data. 10,000 \emph{fBm} paths generated with Davies and Harte algorithm, $\tilde{m}^-=4$, $\tilde{m}^+=32$. Note that for $\mathcal{H}=\{0.1,0.2,0.3,0.4\}$ $(4,32)$ may not be the optimal pair --- the values are shown just for a visual presentation.}
	\label{fig:bias432circ}
\end{figure}

\begin{figure}[b!]
	\centering
		\includegraphics[width=.6\textwidth]{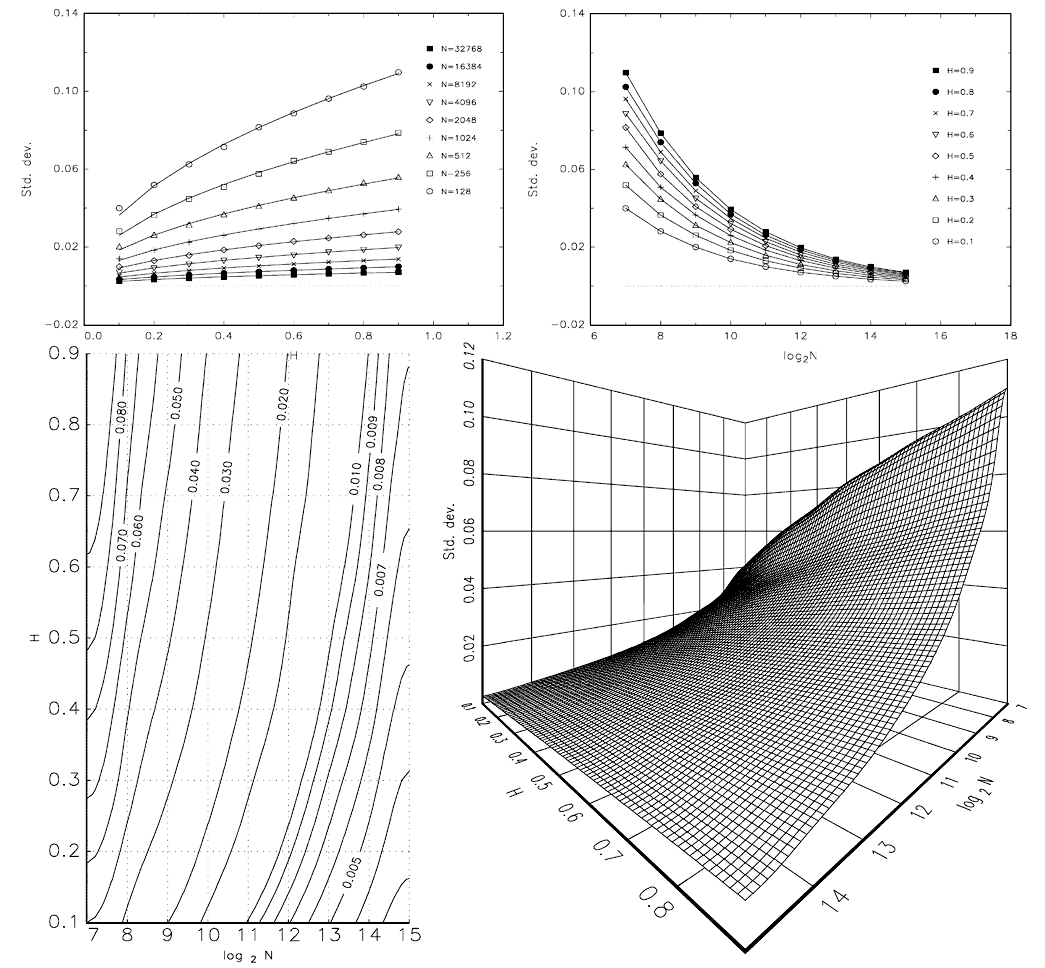}
	\caption{Same as Figure \ref{fig:bias432circ} but instead of bias --- standard deviation.}
	\label{fig:std432circ}
\end{figure}

\begin{figure}[t!]
	\centering
		\includegraphics[width=.52\textwidth]{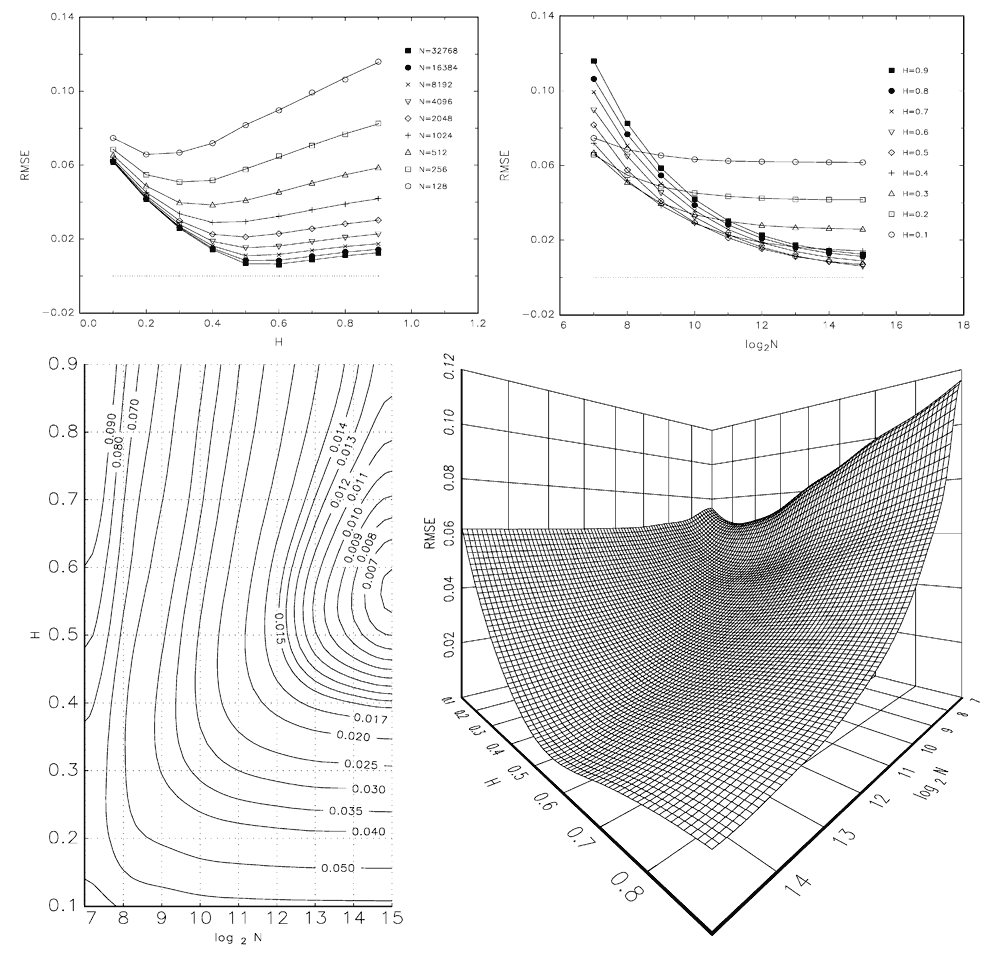}
	\caption{Same as Figure \ref{fig:bias432circ} but instead of bias --- root mean-squared error.}
	\label{fig:rmse432circ}
\end{figure}

$\vartheta(m^-,m^+)$ for different generators for the same $N$ and rank ($\#1, \#2, \#3$) are very close to each another, which indicates that DFA behaved in similar way on the simulated processes. Hence, to picture bias, standard deviation and root mean-squared error of $\hat{H}$ we have chosen Davies and Harte generator of fractional Brownian motion. The generator was one of the fastest and that is why we simulated up to 32,768 observations. Figure~\ref{fig:2p15circ} depicts boxplots of the bias of the estimated Hurst parameter $H=0.1,\ldots,0.9$ on 10,000 \emph{fbm} series of length $N=32768$ for all possible block combinations. In Table \ref{tab:fbm_circ} we listed bias, standard deviation and RMSE for the best three $(\tilde{m}^-,\tilde{m}^+)$. To make it more readable we presented the output on Figures \ref{fig:bias432circ} (bias), \ref{fig:std432circ} (std. dev.) and \ref{fig:rmse432circ} (RMSE) as a function of $H$, $\log_2N$. On lower panels of the Figures \ref{fig:bias432circ}--\ref{fig:rmse432circ} we plot cubic splines with their contours for $H\times \log_2N$. For the purpose of a visual presentation of the behavior of DFA we extended the set of available Hurst parameters to $\mathcal{H}=\{0.1,\ldots,0.9\}$ bearing in mind that $\vartheta(\tilde{m}^-,\tilde{m}^+)$ was computed on $\mathcal{H}=\{0.5,\ldots,0.9\}$.

Let us give a short description of the behavior of the estimator. Let's consider $(\tilde{m}^-,\tilde{m}^+)=(4,32)$. For the series of length $N=2^{10},\ldots,2^{15}$ with nominal $H=1/2$ bias is about $0.004$ -- $0.005$, while for such a span of $N$ standard deviation decreases from $0.029$ to $0.005$ and RMSE decreases from $0.029$ to $0.007$ (Table~\ref{tab:fbm_circ}). With the increase of $H$ from~0.5 to~$0.9$ bias changes from $0.004$ to $-0.014$ for $N=1024$ and from $0.005$ to $-0.010$ for $N=32768$. Standard deviation increases from $0.005$ for $H=1/2$ to $0.007$ for $H=0.9$ for longest available series.
The behavior of DFA for $H=0.1,0.2,0.3,0.4$, which is not presented in tables, can be seen on Figures \ref{fig:bias432circ} -- \ref{fig:rmse432circ}. 

Let us also compare obtained results using the optimal pair of minimal and maximal blocks: $(\tilde{m}^-,\tilde{m}^+)=(4,32)$ and the pairs suggested by Ref. \cite{phA:241:606}: $(8,256)$ for $N=1024$; $(16,256)$ for $N=8192$; $(64,1024)$ for $N=32768$. In our simulation these pairs are respectively: \#8 out of 21, \#13 out of 45 and \#32 out of 66 possible combinations. We have compared bias and standard deviation for such scales in Table \ref{tab:compar1} for $N=1024$ and $N=8192$ and in Table \ref{tab:compar2} for $N=32768$. Let us briefly describe the results. For the series of length $N=1024$ bias for $(\tilde{m}^-,\tilde{m}^+)=(4,32)$ is lower from $1.7$ up to $4.7$ times than in the $(8,256)$ combination and standard deviation is lower about $1.5$ times. With the increase of the length of simulated paths ($N=8192$) the difference in bias is getting smaller --- $4.7$ times bigger for $H=0.6$ and equal for $H=0.9$, but the difference in standard deviation increases to two times. For the longest series available ($N=32768$) bias is lower (about a half) in favor of the $(64,1024)$ combination (except for $H=0.6$) but standard deviation rose almost up to 4~times than in the optimal pair. That is why $\vartheta(64,1024)$ was 32nd out of 66 possible combinations.

\begin{table}[b!]
\caption{The best three block combinations ($\tilde{m}^-,\tilde{m}^+$) and $\vartheta(\tilde{m}^-,\tilde{m}^+)$ for \emph{fBm}/\emph{fGn} series of length $N=2^7,\ldots,2^{15}$ and $H=0.5,0.6,\ldots,0.9$. With $\times$ we marked not simulated --- due to complexity of the algorithm --- series of length $N$. Generators: \texttt{hos} -- Hosking, \texttt{dh} -- Davies and Harte, \texttt{chol} -- Cholesky decomposition, \texttt{pax} -- Paxson, \texttt{ber} -- Beran, \texttt{dl} -- Durbin--Levinson.}
\label{tab:fbmfgn_all}
\small
\begin{tabular}{rrrrrrrrrrrrrrrrrrrrrcccccc}
\hline
$N$&&&\multicolumn{17}{c}{$(m^-,m^+)$}&\multicolumn{6}{c}{$\vartheta(\tilde{m}^-,\tilde{m}^+)$}\\
\cline{4-20}\cline{22-27}
&&&\multicolumn{2}{c}{\texttt{hos}}&&\multicolumn{2}{c}{\texttt{dh}}&&\multicolumn{2}{c}{\texttt{chol}}&&\multicolumn{2}{c}{\texttt{pax}}&&\multicolumn{2}{c}{\texttt{ber}}&&\multicolumn{2}{c}{\texttt{dl}}&&\texttt{hos}&\texttt{dh}&\texttt{chol}&\texttt{pax}&\texttt{ber}&\texttt{dl}\\
\cline{4-5}\cline{7-8}\cline{10-11}\cline{13-14}\cline{16-17}\cline{19-20}\cline{22-27}
128&\#1&&(4,&64)&&(4,&32)&&(4,&32)&&(4,&32)&&(4,&32)&&(4,&32)&&0.0465&0.0493&0.0495&0.0484&0.0481&0.0483\\
&\#2&&(4,&32)&&(4,&64)&&(4,&64)&&(4,&64)&&(4,&64)&&(4,&64)&&0.0479&0.0499&0.0501&0.0493&0.0482&0.0486\\
&\#3&&(4,&128)&&(4,&128)&&(4,&128)&&(4,&128)&&(4,&128)&&(4,&128)&&0.0515&0.0567&0.0560&0.0573&0.0545&0.0558\\
\\
256&\#1&&(4,&64)&&(4,&32)&&(4,&32)&&(4,&32)&&(4,&32)&&(4,&32)&&0.0236&0.0252&0.0248&0.0246&0.0245&0.0247\\
&\#2&&(4,&32)&&(4,&64)&&(4,&64)&&(4,&64)&&(4,&64)&&(4,&64)&&0.0243&0.0254&0.0251&0.0247&0.0250&0.0248\\
&\#3&&(4,&128)&&(4,&128)&&(4,&128)&&(4,&128)&&(4,&128)&&(4,&128)&&0.0261&0.0289&0.0285&0.0287&0.0286&0.0284\\
\\
512&\#1&&(4,&64)&&(4,&32)&&(4,&32)&&(4,&32)&&(4,&32)&&(4,&32)&&0.0122&0.0127&0.0125&0.0124&0.0125&0.0126\\
&\#2&&(4,&32)&&(4,&64)&&(4,&64)&&(4,&64)&&(4,&64)&&(4,&64)&&0.0128&0.0130&0.0127&0.0127&0.0127&0.0129\\
&\#3&&(4,&128)&&(4,&128)&&(4,&128)&&(4,&128)&&(4,&128)&&(4,&128)&&0.0135&0.0146&0.0145&0.0146&0.0145&0.0146\\
\\
1024&\#1&&(4,&64)&&(4,&32)&&(4,&32)&&(4,&32)&&(4,&32)&&(4,&32)&&0.0066&0.0065&0.0065&0.0064&0.0064&0.0064\\
&\#2&&(4,&32)&&(4,&64)&&(4,&64)&&(4,&64)&&(4,&64)&&(4,&64)&&0.0068&0.0067&0.0067&0.0065&0.0066&0.0065\\
&\#3&&(4,&128)&&(4,&128)&&(4,&128)&&(4,&128)&&(4,&128)&&(4,&128)&&0.0071&0.0077&0.0075&0.0075&0.0075&0.0074\\
\\
2048&\#1&&(4,&64)&&(4,&32)&&(4,&32)&&(4,&32)&&(4,&32)&&(4,&32)&&0.0036&0.0033&0.0033&0.0033&0.0033&0.0033\\
&\#2&&(4,&32)&&(4,&64)&&(4,&64)&&(4,&64)&&(4,&64)&&(4,&64)&&0.0037&0.0035&0.0034&0.0034&0.0034&0.0035\\
&\#3&&(4,&128)&&(4,&128)&&(4,&128)&&(4,&128)&&(4,&128)&&(4,&128)&&0.0039&0.0040&0.0039&0.0039&0.0039&0.0039\\
\\
4096&\#1&&(4,&64)&&(4,&32)&&(4,&32)&&(4,&32)&&(4,&32)&&(4,&32)&&0.0022&0.0018&0.0018&0.0018&0.0018&0.0018\\
&\#2&&(4,&32)&&(4,&64)&&(4,&64)&&(4,&64)&&(4,&64)&&(4,&64)&&0.0023&0.0019&0.0019&0.0019&0.0019&0.0019\\
&\#3&&(4,&128)&&(4,&128)&&(4,&128)&&(4,&128)&&(4,&128)&&(4,&128)&&0.0023&0.0021&0.0021&0.0021&0.0021&0.0021\\
\\
8192&\#1&&(4,&64)&&(4,&32)&&&&&(4,&32)&&&&&&&&0.0014&0.0010&&0.0010&&\\
&\#2&&(4,&128)&&(4,&64)&&\multicolumn{2}{c}{$\times$}&&(4,&64)&&\multicolumn{2}{c}{$\times$}&&\multicolumn{2}{c}{$\times$}&&0.0015&0.0011&$\times$&0.0011&$\times$&$\times$\\
&\#3&&(4,&32)&&(4,&128)&&&&&(4,&128)&&&&&&&&0.0015&0.0013&&0.0012&&\\
\\
16384&\#1&&&&&(4,&32)&&&&&(4,&32)&&&&&&&&&0.0006&&0.0006&&\\
&\#2&&\multicolumn{2}{c}{$\times$}&&(4,&64)&&\multicolumn{2}{c}{$\times$}&&(4,&64)&&\multicolumn{2}{c}{$\times$}&&\multicolumn{2}{c}{$\times$}&&$\times$&0.0007&$\times$&0.0007&$\times$&$\times$\\
&\#3&&&&&(4,&128)&&&&&(4,&128)&&&&&&&&&0.0008&&0.0008&&\\
\\
32768&\#1&&&&&(4,&32)&&&&&&&&&&&&&&&0.0004&&&&\\
&\#2&&\multicolumn{2}{c}{$\times$}&&(4,&64)&&\multicolumn{2}{c}{$\times$}&&\multicolumn{2}{c}{$\times$}&&\multicolumn{2}{c}{$\times$}&&\multicolumn{2}{c}{$\times$}&&$\times$&0.0005&$\times$&$\times$&$\times$&$\times$\\
&\#3&&&&&(4,&128)&&&&&&&&&&&&&&&0.0006&&&&\\
\hline
\end{tabular}
\end{table} 
\normalsize

\clearpage

\begin{figure}[b!]
	\centering
		\includegraphics[width=.79\textwidth]{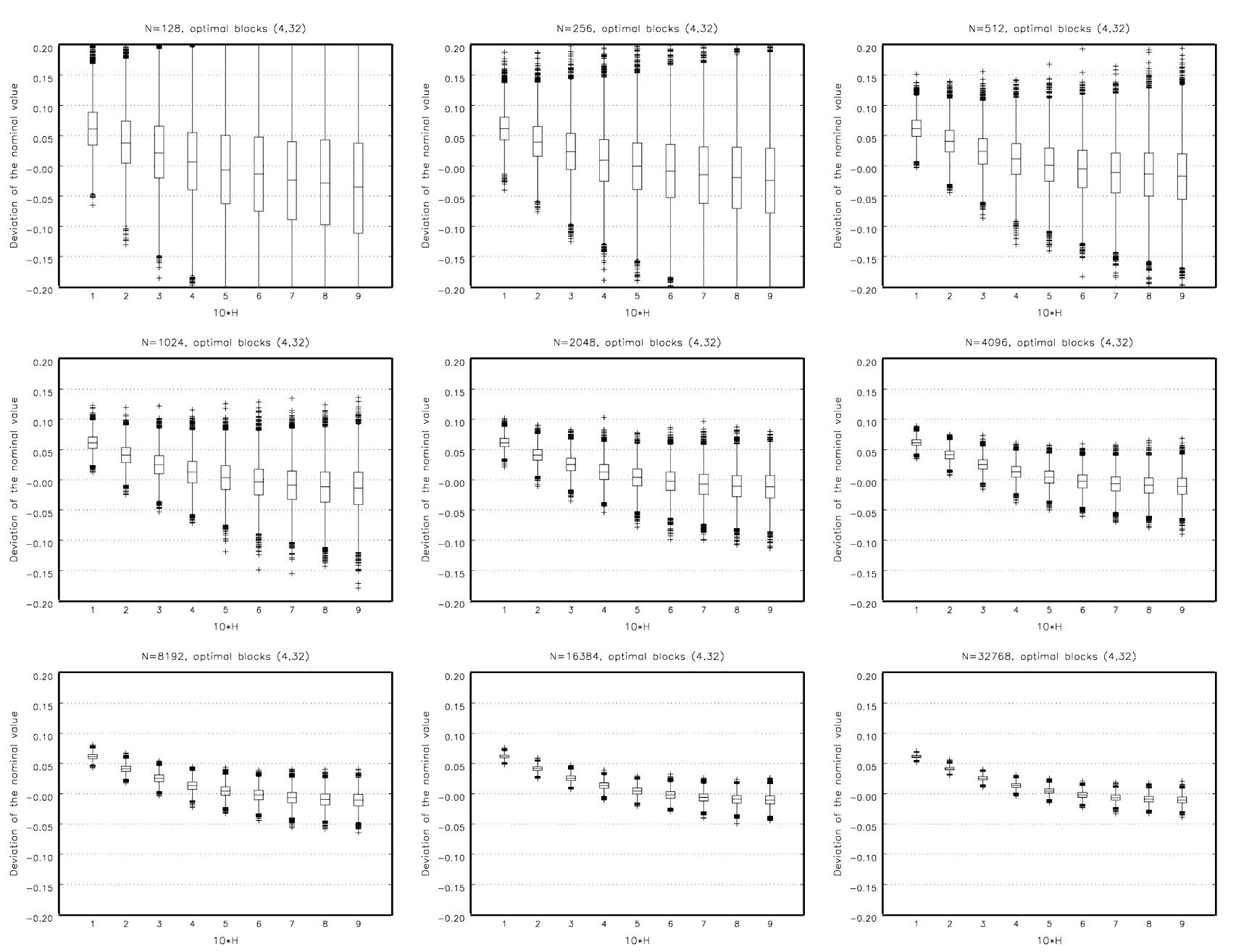}
	\caption{The optimal pair of minimal and maximal blocks $(\tilde{m}^-,\tilde{m}^+)=(4,32)$ and boxplots of the bias of the estimated Hurst parameter $H=0.1,\ldots,0.9$ for 10,000 \emph{fBm} series of length $N=2^7,\ldots,2^{15}$ simulated by Davies and Harte method. Note: for $H=0.1,\ldots,0.4$ the pair (4,32) may not be optimal.}
	\label{fig:boxpall432}
\end{figure}

\clearpage

\clearpage

\begin{table}[b!]
\caption{Bias, standard deviation and root mean square error for the best three block combinations. 10,000 estimates of $H$ for different lengths of $N$, minimal $m^-$ and maximal $m^+$ blocks. Process: \fbm, generator: Davies and Harte.}
\label{tab:fbm_circ}
\footnotesize
\[\begin{array}{rrrrrcrrrrrrrrrrrrrrrrrrr}
\hline
N&&(\tilde{m}^-,&\tilde{m}^+)&&\vartheta&&\multicolumn{5}{c}{\text{Bias}}&&\multicolumn{5}{c}{\text{Std. dev.}}&&\multicolumn{5}{c}{\text{RMSE}}\\
\cline{8-12}\cline{14-18}\cline{20-24}
\cline{8-12}\cline{14-18}\cline{20-24}
&&&&&&&0.5&0.6&0.7&0.8&0.9&&0.5&0.6&0.7&0.8&0.9&&0.5&0.6&0.7&0.8&0.9\\
\cline{1-6}\cline{8-12}\cline{14-18}\cline{20-24}
128&\#1&(4,&32)&&0.0493&&-0.005&-0.013&-0.024&-0.028&-0.037&&0.082&0.089&0.096&0.102&0.110&&0.082&0.090&0.099&0.106&0.116\\
&\#2&(4,&64)&&0.0499&&-0.016&-0.023&-0.034&-0.039&-0.048&&0.078&0.087&0.093&0.101&0.107&&0.080&0.090&0.099&0.108&0.118\\
&\#3&(4,&128)&&0.0567&&-0.025&-0.033&-0.044&-0.051&-0.059&&0.080&0.089&0.096&0.105&0.111&&0.084&0.095&0.105&0.117&0.126\\
256&\#1&(4,&32)&&0.0252&&0.000&-0.009&-0.015&-0.020&-0.025&&0.058&0.064&0.069&0.074&0.079&&0.058&0.065&0.071&0.077&0.083\\
&\#2&(4,&64)&&0.0254&&-0.009&-0.015&-0.022&-0.027&-0.031&&0.056&0.063&0.067&0.072&0.077&&0.057&0.065&0.071&0.077&0.083\\
&\#3&(4,&128)&&0.0289&&-0.014&-0.022&-0.027&-0.034&-0.038&&0.058&0.065&0.071&0.076&0.081&&0.060&0.069&0.076&0.083&0.089\\
512&\#1&(4,&32)&&0.0127&&0.002&-0.005&-0.011&-0.014&-0.018&&0.041&0.045&0.049&0.053&0.056&&0.041&0.045&0.050&0.055&0.059\\
&\#2&(4,&64)&&0.0130&&-0.004&-0.010&-0.015&-0.019&-0.022&&0.040&0.045&0.049&0.053&0.055&&0.040&0.046&0.051&0.056&0.060\\
&\#3&(4,&128)&&0.0146&&-0.009&-0.014&-0.018&-0.022&-0.026&&0.042&0.047&0.051&0.055&0.058&&0.043&0.049&0.054&0.059&0.064\\
1024&\#1&(4,&32)&&0.0065&&0.004&-0.004&-0.009&-0.012&-0.014&&0.029&0.032&0.035&0.037&0.039&&0.029&0.032&0.036&0.039&0.042\\
&\#2&(4,&64)&&0.0067&&-0.003&-0.008&-0.013&-0.015&-0.016&&0.029&0.032&0.035&0.037&0.040&&0.029&0.033&0.037&0.040&0.043\\
&\#3&(4,&128)&&0.0077&&-0.006&-0.010&-0.014&-0.016&-0.018&&0.030&0.034&0.037&0.039&0.042&&0.031&0.035&0.039&0.042&0.046\\
2048&\#1&(4,&32)&&0.0033&&0.004&-0.002&-0.007&-0.011&-0.012&&0.021&0.023&0.025&0.026&0.028&&0.021&0.023&0.026&0.028&0.030\\
&\#2&(4,&64)&&0.0035&&-0.001&-0.007&-0.010&-0.013&-0.014&&0.021&0.023&0.024&0.026&0.028&&0.021&0.024&0.027&0.029&0.031\\
&\#3&(4,&128)&&0.0040&&-0.004&-0.009&-0.012&-0.013&-0.015&&0.022&0.024&0.026&0.028&0.029&&0.022&0.025&0.028&0.031&0.033\\
4096&\#1&(4,&32)&&0.0018&&0.005&-0.003&-0.007&-0.009&-0.011&&0.015&0.016&0.017&0.019&0.020&&0.015&0.016&0.019&0.021&0.023\\
&\#2&(4,&64)&&0.0019&&-0.001&-0.007&-0.009&-0.011&-0.013&&0.014&0.016&0.017&0.019&0.020&&0.014&0.017&0.020&0.022&0.023\\
&\#3&(4,&128)&&0.0021&&-0.004&-0.008&-0.010&-0.012&-0.013&&0.015&0.017&0.018&0.020&0.021&&0.016&0.019&0.021&0.023&0.024\\
8192&\#1&(4,&32)&&0.0010&&0.004&-0.002&-0.006&-0.009&-0.011&&0.010&0.011&0.012&0.013&0.014&&0.011&0.011&0.014&0.016&0.017\\
&\#2&(4,&64)&&0.0011&&-0.001&-0.006&-0.009&-0.011&-0.012&&0.010&0.011&0.012&0.013&0.014&&0.010&0.013&0.015&0.017&0.018\\
&\#3&(4,&128)&&0.0013&&-0.003&-0.007&-0.010&-0.011&-0.012&&0.011&0.012&0.013&0.014&0.015&&0.011&0.014&0.016&0.018&0.019\\
16384&\#1&(4,&32)&&0.0006&&0.004&-0.002&-0.006&-0.009&-0.010&&0.007&0.008&0.009&0.009&0.010&&0.009&0.008&0.011&0.013&0.014\\
&\#2&(4,&64)&&0.0007&&-0.001&-0.006&-0.009&-0.011&-0.012&&0.007&0.008&0.009&0.009&0.010&&0.007&0.010&0.012&0.014&0.015\\
&\#3&(4,&128)&&0.0008&&-0.003&-0.007&-0.009&-0.011&-0.011&&0.008&0.008&0.009&0.010&0.011&&0.008&0.011&0.013&0.015&0.016\\
32768&\#1&(4,&32)&&0.0004&&0.005&-0.002&-0.006&-0.009&-0.010&&0.005&0.006&0.006&0.007&0.007&&0.007&0.006&0.009&0.011&0.012\\
&\#2&(4,&64)&&0.0005&&-0.001&-0.006&-0.009&-0.011&-0.012&&0.005&0.006&0.006&0.007&0.007&&0.005&0.008&0.011&0.013&0.014\\
&\#3&(4,&128)&&0.0006&&-0.003&-0.007&-0.009&-0.011&-0.011&&0.005&0.006&0.007&0.007&0.007&&0.006&0.009&0.011&0.013&0.013\\
\hline
\end{array}\]
\end{table} 
\normalsize

\clearpage

\begin{figure}[t!]
	\centering
		\includegraphics[width=1\textwidth]{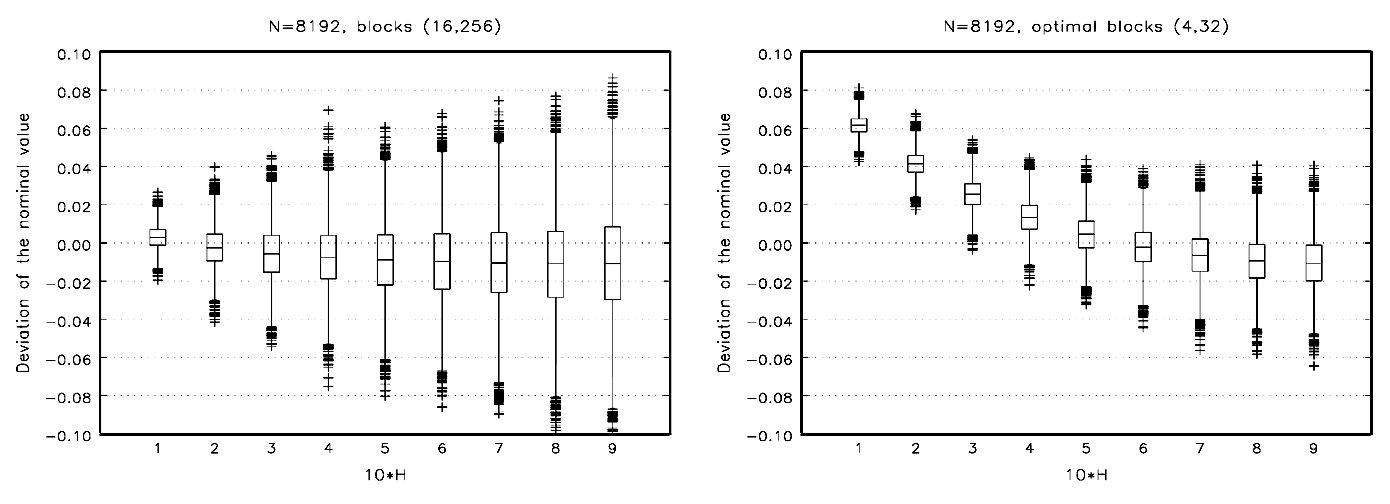}
	\caption{Comparison of boxplots of the bias of the estimated Hurst parameter for the the optimal pair of minimal and maximal blocks~$(4,32)$ and for~$(64,1024)$.  10,000 paths of length $N=8192$ of \emph{fBm} simulated by Davies and Harte method.}
	\label{fig:boxcompare}
\end{figure}
\begin{table}[t!]
\caption{Comparison of bias and standard deviation for the $(\tilde{m}^-,\tilde{m}^+)=(4,32)$ and $(8,256)$, $(16,256)$ block combinations for the series of $N=1024$ and $N=8192$ respectively.}
\label{tab:compar1}
\small
\[\begin{array}{rrrrrrrrrrrrrrrrr}
\hline
H&&\multicolumn{7}{l}{N=1024}&&\multicolumn{7}{l}{N=8192}\\
\cline{3-9}\cline{11-17}
&&\multicolumn{2}{l}{\text{Bias}}&&&\multicolumn{2}{l}{\text{Std. dev.}}&&&\multicolumn{2}{l}{\text{Bias}}&&&\multicolumn{2}{l}{\text{Std. dev.}}&\\
\cline{3-5}\cline{7-9}\cline{11-13}\cline{15-17}
&&(8,256)&(4,32)&&&(8,256)&(4,32)&&&(16,256)&(4,32)&&&(16,256)&(4,32)&\\
&&(a)&(b)&(a)/(b)&&(a)&(b)&(a)/(b)&&(a)&(b)&(a)/(b)&&(a)&(b)&(a)/(b)\\
\hline
0.5&&-0.016&0.004&-4.4&&0.042&0.029&1.4&&-0.009&0.004&-2.0&&0.020&0.010&1.9\\
0.6&&-0.017&-0.004&4.7&&0.046&0.032&1.4&&-0.010&-0.002&4.7&&0.022&0.011&1.9\\
0.7&&-0.021&-0.009&2.3&&0.051&0.035&1.5&&-0.011&-0.006&1.6&&0.024&0.012&1.9\\
0.8&&-0.022&-0.012&1.9&&0.054&0.037&1.5&&-0.011&-0.009&1.2&&0.026&0.013&2.0\\
0.9&&-0.025&-0.014&1.7&&0.058&0.039&1.5&&-0.011&-0.011&1.0&&0.028&0.014&2.0\\
\hline
\multicolumn{16}{l}{\text{Note: $(a)$, $(b)$, $(a/b)$ are rounded numbers.}}
\end{array}\]
\end{table}
\normalsize
\begin{table}[t!]
\caption{Comparison of bias and standard deviation for the $(4,32)$ and $(64,1024)$ block combinations for the series of length~$N=32768$.}
\label{tab:compar2}
\small
\[\begin{array}{rrrrrrrrrr}
\hline
H&&\multicolumn{7}{l}{N=32768}\\
\cline{3-9}
&&\multicolumn{2}{l}{\text{Bias}}&&&\multicolumn{2}{l}{\text{Std. dev.}}&\\
\cline{3-5}\cline{7-9}
&&(64,1024)&(4,32)&&&(64,1024)&(4,32)&\\
&&(a)&(b)&(a/b)&&(a)&(b)&(a/b)\\
\hline
0.5&&-0.003&0.005&-0.7&&0.019&0.005&3.7\\
0.6&&-0.003&-0.002&1.8&&0.022&0.006&3.8\\
0.7&&-0.004&-0.006&0.6&&0.024&0.006&3.8\\
0.8&&-0.004&-0.009&0.5&&0.026&0.007&3.8\\
0.9&&-0.004&-0.010&0.4&&0.027&0.007&3.9\\
\hline
\end{array}\]
\end{table}
\normalsize

\subsection{The polynomial trend fit}\label{subsec:poly}

Tables and figures for this and the following subsection were moved to appendix \ref{appx}.

We have also tried to investigate the impact of different orders of polynomial fit on the properties of DFA. Hence, we have chosen linear, quadratic, cubic and fourth order polynomial fit. The results are presented in Table \ref{tab:fbm_circ_poly}. Our optimization criterion $\vartheta$ varies slightly for long series. One can use 4th order polynomial fit for the series of length of $N=32768$ and $(\tilde{m}^-,\tilde{m}^+)=(16,256)$ where bias is $0.006$ for $H=1/2$ and $0.000$ for $H=0.8,0.9$ with standard deviation only approximately $0.002$ grater than in the case of using the linear fit, but should not use $r=4$ for e.g. N=1024, where bias for $H=1/2$ is $0.026$ comparing to 0.004 for $r=1$ --- six time higher. For the simplicity of empirical analysis we suggest to use the simplest case which is linear trend fit. Note that $(\tilde{m}^-,\tilde{m}^+)$ is the same for different $N$ only for that kind of fit, what is more convenient to carry out empirical analysis. 

\subsection{Impact of short memory --- \textup{ARFIMA}(p,d,q)}\label{subsec:arfima}

We have examined DFA on the simulated ARIFMA($p,d,q$) series in the separate subsection due to the following reasons
\begin{itemize}
	\item The asymptotic expansion of ARIFMA(0,d,0) processes, which may influence final results \cite{taqqu98}.
	\item Ability to verify the impact of additional short-range correlations, i.e., $p,q>0$, on the blocks adjustment.
\end{itemize}

We have simulated ARFIMA$(p,d,q)$ in the frequency domain using fast Fourier transform based on S-PLUS code written originally by \cite{beran94} and then checked sensitivity of final results on short-range dependence for ARFIMA$(p,d,q)$ series: $(0,d,1)$ with $\theta=0.5$, $(1,d,0)$ with $\phi=0.5$, $(1,d,1)$ with $\phi=0.3,\theta=0.7$, $(1,d,1)$ with $\phi=-0.3,\theta=-0.7$ and $(1,d,1)$ with $\phi=0.7,\theta=0.3$. Summary results for the best three block combinations are in Table \ref{tab:arfima_all}.

Let us describe obtained results. DFA behaves differently on autoregressive fractional integrated moving average process than on fractional Gaussian noise or fractional Brownian motion. For ARFIMA($0,d,0$) with up to $N=512$ observations, among the best three block combinations there is $(4,32)$ (the pair $(4,64)$ minimizes $\vartheta$). In Table~\ref{tab:dh_arfima0d0} we see that --- contrary to bias --- standard deviation for different Hurst parameters is almost identical. Except for $H=0.5$, bias is much larger for ARFIMA$(0,d,0)$ than for Davies and Harte generator and that is why $(\tilde{m}^-,\tilde{m}^+)=(4,32)$ is no longer valid for such process.

In Table~\ref{tab:dh_arfimapdq} we have presented the sensitivity of DFA on the different values of autoregressive~$\phi$ and moving average~$\theta$ parameters for the same block combination $(4,32)$. Introducing MA part leads to strong negative bias (see Figure~\ref{fig:arfima0d1}) while AR causes strong positive bias (Figure~\ref{fig:arfima1d0}) (variance of the estimator remains stable). This results in the different than $(4,32)$ optimal pairs listed in Table~\ref{tab:dh_arfimapdq2}. Although we managed to decrease the bias, standard deviation rose, which makes DFA incapable to estimate Hurst parameter precisely in terms of the presence of short memory.

DFA ''prefers'' much larger scales $(\tilde{m}^-,\tilde{m}^+)$ under the presence of short memory, but for the same length of the series these scales are different (Table~\ref{tab:arfima_all}). After the joint introduction of AR and MA parts we still observe strong bias and steady variance. For $\phi=0.7>\theta=0.3$ bias is positive and if $\phi=0.3<\theta=0.7$ --- negative (Figures \ref{fig:arfima1d1a}, \ref{fig:arfima1d1c}). Negative $\phi=-0.3$ and $\theta=-0.7$ parameters (Figure~\ref{fig:arfima1d1b}) lead to greater bias than in the case of lack of short memory, but the effect is not as strong as for positive values of $\phi$ and $\theta$ (and smaller scales are preferable).

Due to the complexity of the algorithm, we have restricted that part of our analysis to six different $(p,d,q)$ only (three for $(1,d,1)$). To obtain the full picture of the behavior of DFA on ARFIMA$(0,d,1)$, $(1,d,0)$ and $(1,d,1)$ we should carry out our simulations for $\phi,\theta=0,\pm .1,\ldots,\pm 0.9$, $d=0,\pm0.1,\ldots,\pm0.4$ and $N=2^7,\ldots,2^p$. It took about 24 hours to simulate 10,000 certain ARFIMA$(1,d,1)$ series of length $N=8192$ with nine different values of $d$ parameter. Hence, such extended and precise analysis (in terms of the number of replications) would be very time-consuming.  Nevertheless we have tried to find best block combinations on available six simulated ARFIMA$(p,d,q)$ processes, searching for the pair $(\tilde{m}^-,\tilde{m}^+)$ that minimizes the sum of MSE for $H=0.5,\ldots,0.9$ and these six processes altogether, bearing in mind that the results will be preliminary. We have listed them in Table~\ref{tab:dh_arfimapdq3}. Although we found the combination for the series of length $N$, we see that the variance of DFA is too large (standard deviation for 8192 observations is about $0.04$--$0.06$ with bias varying form $-0.09$ to $0.029$).

\section{Conclusions}

Blocks adjustment may significantly improve precision of DFA for persistent processes. These improvements are presented in Tables \ref{tab:compar1}, \ref{tab:compar2} and depicted on Figures~\ref{fig:2p15circ},~\ref{fig:boxpall432} and~\ref{fig:boxcompare}. For the sake of simplicity of analysis we recommend using linear trend fit (Table~\ref{tab:fbm_circ_poly}). Through extensive simulations we have shown that the optimal pair of minimal and maximal blocks is $(4,32)$, which may reduce standard deviation even up to four times for the series of length $N=32768$. Second and third results are pairs $(4,64)$, $(4,128)$ respectively. The results are robust on the length of the series and the type of generators of fractional Brownian motion or fractional Gaussian noise. The exception is ARFIMA$(0,d,0)$ --- blocks adjustment also improved the quality of the estimator, but indicated different block combinations as optimal (Table~\ref{tab:arfima_all}). 

DFA is very sensitive to the presence of short-range correlations in the series. The bias and variance are reduced at the expense of the other --- exclusion of large-sized blocks reduces variance but results in greater bias, contrary to cuts of small blocks --- but the effect is so strong that it disables precise estimation of the long memory parameter. In such cases data must be filtered or DFA should be replaced e.g., with  Global log-periodogram \cite{doukhan03}.

\section{Acknowledgments}

The author would like to express his deepest gratitude to Professor S\l{}awomir Dorosiewicz for his insights and his support in helping him give this work its final form.


\appendix

\section{Tables and figures for sections \ref{subsec:poly} and \ref{subsec:arfima}}
\label{appx}

\begin{table}[b!]
\caption{Impact of order $r=1,2,\ldots,4$ of the polynomial trend fit on bias, standard deviation and root mean-squared error for the best block combination $(\tilde{m}^-,\tilde{m}^+)$. Generator: Davies and Harte method for 10,000 paths of fractional Brownian motion.}
\label{tab:fbm_circ_poly}
\footnotesize
\[\begin{array}{rrrrrcrrrrrrrrrrrrrrrrrrr}
\hline
N&r&(\tilde{m}^-,&\tilde{m}^+)&&\vartheta&&\multicolumn{5}{c}{\text{Bias}}&&\multicolumn{5}{c}{\text{Std. dev.}}&&\multicolumn{5}{c}{\text{RMSE}}\\
\cline{8-12}\cline{14-18}\cline{20-24}
\cline{8-12}\cline{14-18}\cline{20-24}
&&&&&&&0.5&0.6&0.7&0.8&0.9&&0.5&0.6&0.7&0.8&0.9&&0.5&0.6&0.7&0.8&0.9\\
\cline{1-6}\cline{8-12}\cline{14-18}\cline{20-24}
128&1&(4,&32)&&0.0493&&-0.005&-0.013&-0.024&-0.028&-0.037&&0.082&0.089&0.096&0.102&0.110&&0.082&0.090&0.099&0.106&0.116\\
&2&(4,&128)&&0.0398&&0.045&0.036&0.024&0.018&0.008&&0.072&0.079&0.083&0.091&0.095&&0.085&0.086&0.086&0.092&0.096\\
&3&(8,&128)&&0.0477&&0.010&0.002&-0.009&-0.012&-0.023&&0.081&0.090&0.096&0.104&0.111&&0.081&0.090&0.096&0.104&0.114\\
&4&(8,&128)&&0.0456&&0.046&0.037&0.025&0.021&0.012&&0.076&0.084&0.089&0.097&0.103&&0.089&0.091&0.093&0.099&0.104\\
256&1&(4,&32)&&0.0252&&0.000&-0.009&-0.015&-0.020&-0.025&&0.058&0.064&0.069&0.074&0.079&&0.058&0.065&0.071&0.077&0.083\\
&2&(4,&256)&&0.0227&&0.032&0.023&0.015&0.008&0.003&&0.054&0.060&0.065&0.069&0.073&&0.063&0.064&0.067&0.069&0.073\\
&3&(8,&128)&&0.0250&&0.017&0.009&0.003&-0.003&-0.006&&0.058&0.065&0.070&0.075&0.081&&0.060&0.065&0.070&0.075&0.081\\
&4&(8,&256)&&0.0237&&0.032&0.024&0.016&0.010&0.005&&0.055&0.061&0.066&0.071&0.075&&0.063&0.066&0.068&0.071&0.075\\
512&1&(4,&32)&&0.0127&&0.002&-0.005&-0.011&-0.014&-0.018&&0.041&0.045&0.049&0.053&0.056&&0.041&0.045&0.050&0.055&0.059\\
&2&(4,&512)&&0.0141&&0.023&0.016&0.011&0.006&0.000&&0.043&0.047&0.051&0.055&0.058&&0.049&0.050&0.053&0.055&0.058\\
&3&(8,&128)&&0.0135&&0.020&0.014&0.009&0.006&0.002&&0.041&0.047&0.050&0.054&0.058&&0.046&0.049&0.051&0.055&0.058\\
&4&(8,&512)&&0.0140&&0.023&0.016&0.011&0.007&0.002&&0.042&0.047&0.052&0.055&0.058&&0.048&0.049&0.053&0.055&0.058\\
1024&1&(4,&32)&&0.0065&&0.004&-0.004&-0.009&-0.012&-0.014&&0.029&0.032&0.035&0.037&0.039&&0.029&0.032&0.036&0.039&0.042\\
&2&(8,&128)&&0.0084&&0.000&-0.004&-0.008&-0.009&-0.011&&0.033&0.037&0.041&0.043&0.046&&0.033&0.037&0.041&0.044&0.047\\
&3&(8,&256)&&0.0072&&0.013&0.009&0.005&0.002&0.000&&0.031&0.034&0.038&0.040&0.043&&0.034&0.036&0.038&0.040&0.043\\
&4&(8,&512)&&0.0085&&0.026&0.021&0.017&0.013&0.011&&0.031&0.034&0.038&0.039&0.042&&0.041&0.040&0.041&0.041&0.044\\
2048&1&(4,&32)&&0.0033&&0.004&-0.002&-0.007&-0.011&-0.012&&0.021&0.023&0.025&0.026&0.028&&0.021&0.023&0.026&0.028&0.030\\
&2&(8,&128)&&0.0042&&0.002&-0.003&-0.005&-0.007&-0.008&&0.024&0.026&0.028&0.031&0.032&&0.024&0.026&0.029&0.032&0.033\\
&3&(8,&256)&&0.0040&&0.016&0.010&0.007&0.005&0.004&&0.022&0.024&0.027&0.029&0.030&&0.027&0.027&0.028&0.029&0.030\\
&4&(16,&256)&&0.0051&&0.005&0.001&-0.001&-0.003&-0.003&&0.026&0.029&0.032&0.035&0.036&&0.027&0.029&0.032&0.035&0.036\\
4096&1&(4,&32)&&0.0018&&0.005&-0.003&-0.007&-0.009&-0.011&&0.015&0.016&0.017&0.019&0.020&&0.015&0.016&0.019&0.021&0.023\\
&2&(8,&128)&&0.0021&&0.002&-0.002&-0.004&-0.005&-0.007&&0.017&0.018&0.020&0.022&0.023&&0.017&0.019&0.021&0.022&0.024\\
&3&(8,&256)&&0.0023&&0.016&0.011&0.009&0.007&0.005&&0.016&0.017&0.019&0.020&0.021&&0.022&0.021&0.021&0.022&0.022\\
&4&(16,&256)&&0.0026&&0.005&0.002&0.001&0.000&-0.002&&0.019&0.021&0.023&0.024&0.026&&0.019&0.021&0.023&0.024&0.026\\
8192&1&(4,&32)&&0.0010&&0.004&-0.002&-0.006&-0.009&-0.011&&0.010&0.011&0.012&0.013&0.014&&0.011&0.011&0.014&0.016&0.017\\
&2&(8,&128)&&0.0011&&0.002&-0.001&-0.004&-0.005&-0.006&&0.012&0.013&0.014&0.015&0.016&&0.012&0.013&0.015&0.016&0.017\\
&3&(8,&512)&&0.0013&&0.011&0.008&0.006&0.004&0.003&&0.012&0.013&0.014&0.016&0.017&&0.017&0.016&0.016&0.016&0.017\\
&4&(16,&256)&&0.0013&&0.006&0.003&0.001&0.000&-0.001&&0.013&0.015&0.016&0.017&0.019&&0.014&0.015&0.016&0.017&0.019\\
16384&1&(4,&32)&&0.0006&&0.004&-0.002&-0.006&-0.009&-0.010&&0.007&0.008&0.009&0.009&0.010&&0.009&0.008&0.011&0.013&0.014\\
&2&(8,&128)&&0.0006&&0.002&-0.001&-0.003&-0.005&-0.005&&0.008&0.009&0.010&0.011&0.012&&0.009&0.009&0.011&0.012&0.013\\
&3&(8,&1024)&&0.0008&&0.008&0.006&0.004&0.003&0.002&&0.010&0.011&0.012&0.013&0.014&&0.013&0.012&0.013&0.013&0.014\\
&4&(16,&256)&&0.0007&&0.006&0.003&0.001&0.000&0.000&&0.009&0.011&0.011&0.012&0.013&&0.011&0.011&0.012&0.012&0.013\\
32768&1&(4,&32)&&0.0004&&0.005&-0.002&-0.006&-0.009&-0.010&&0.005&0.006&0.006&0.007&0.007&&0.007&0.006&0.009&0.011&0.012\\
&2&(8,&128)&&0.0003&&0.003&-0.001&-0.003&-0.004&-0.005&&0.006&0.007&0.007&0.008&0.008&&0.006&0.007&0.008&0.009&0.010\\
&3&(16,&256)&&0.0004&&0.000&-0.002&-0.003&-0.004&-0.004&&0.007&0.008&0.009&0.010&0.010&&0.007&0.008&0.010&0.010&0.011\\
&4&(16,&256)&&0.0004&&0.006&0.003&0.001&0.000&0.000&&0.007&0.007&0.008&0.009&0.009&&0.009&0.008&0.008&0.009&0.009\\
\hline
\end{array}\]
\end{table}
\normalsize

\clearpage

\begin{table}[h!]
\caption{The best three block combinations $(\tilde{m}^-,\tilde{m}^+)$ for 10,000 ARFIMA$(p,d,q)$ series of length $N=2^7,\ldots,2^{13}$ and $H=0.5,\ldots,0.9$.}
\label{tab:arfima_all}
\begin{tabular}{rrrrrrrrrrrrrrrrrrrr}
\hline
$N$&&&\multicolumn{17}{c}{$(\tilde{m}^-,\tilde{m}^+)$}\\
\cline{4-20}
&&&\multicolumn{2}{c}{$(0,d,0)$}&&\multicolumn{2}{c}{$(0,d,1)$}&&\multicolumn{2}{c}{$(1,d,0)$}&&\multicolumn{2}{c}{$(1,d,1)$}&&\multicolumn{2}{c}{$(1,d,1)$}&&\multicolumn{2}{c}{$(1,d,1)$}\\
&&&&&&\multicolumn{2}{c}{$\theta=0.5$}&&\multicolumn{2}{c}{$\phi=0.5$}&&\multicolumn{2}{c}{$\phi=0.3$}&&\multicolumn{2}{c}{$\phi=-0.3$}&&\multicolumn{2}{c}{$\phi=0.7$}\\
&&&&&&&&&&&&\multicolumn{2}{c}{$\theta=0.7$}&&\multicolumn{2}{c}{$\theta=-0.7$}&&\multicolumn{2}{c}{$\theta=0.3$}\\
\cline{1-2}\cline{4-5}\cline{7-8}\cline{10-11}\cline{13-14}\cline{16-17}\cline{19-20}
128&\#1&&(4,&64)&&(16,&128)&&(8,&12)&&(16,&128)&&(4,&128)&&(8,&128)\\
&\#2&&(4,&32)&&(8,&128)&&(4,&128)&&(4,&128)&&(4,&64)&&(16,&128)\\
&\#3&&(4,&128)&&(4,&128)&&(16,&128)&&(8,&128)&&(4,&32)&&(4,&128)\\
\\
256&\#1&&(4,&64)&&(16,&256)&&(8,&256)&&(32,&256)&&(4,&128)&&(16,&256)\\
&\#2&&(4,&32)&&(32,&256)&&(16,&256)&&(16,&256)&&(4,&64)&&(8,&256)\\
&\#3&&(4,&128)&&(16,&128)&&(8,&128)&&(8,&256)&&(4,&256)&&(32,&256)\\
\\
512&\#1&&(4,&64)&&(32,&512)&&(8,&512)&&(32,&512)&&(4,&128)&&(16,&512)\\
&\#2&&(4,&128)&&(16,&512)&&(16,&512)&&(64,&512)&&(4,&256)&&(32,&512)\\
&\#3&&(4,&32)&&(32,&256)&&(16,&256)&&(32,&256)&&(4,&64)&&(8,&512)\\
\\
1024&\#1&&(4,&128)&&(32,&1024)&&(8,&1024)&&(64,&1024)&&(4,&256)&&(16,&1024)\\
&\#2&&(4,&256)&&(32,&512)&&(16,&512)&&(32,&1024)&&(4,&512)&&(32,&1024)\\
&\#3&&(4,&64)&&(16,&1024)&&(16,&1024)&&(64,&512)&&(4,&128)&&(32,&512)\\
\\
2048&\#1&&(4,&256)&&(32,&1024)&&(16,&1024)&&(64,&1024)&&(4,&512)&&(16,&2048)\\
&\#2&&(4,&128)&&(32,&2048)&&(8,&2048)&&(64,&2048)&&(8,&128)&&(32,&1024)\\
&\#3&&(4,&512)&&(64,&1024)&&(16,&2048)&&(32,&2048)&&(4,&256)&&(32,&2048)\\
\\
4096&\#1&&(4,&512)&&(64,&1024)&&(16,&2048)&&(64,&2048)&&(8,&128)&&(32,&2048)\\
&\#2&&(4,&256)&&(64,&2048)&&(16,&1024)&&(64,&4096)&&(8,&256)&&(16,&4096)\\
&\#3&&(8,&256)&&(32,&2048)&&(16,&4096)&&(64,&1024)&&(8,&64)&&(32,&4096)\\
\\
8192&\#1&&(8,&512)&&(64,&2048)&&(32,&1024)&&(128&2048)&&(8,&256)&&(32,&4096)\\
&\#2&&(4,&512)&&(64,&1024)&&(16,&4096)&&(64,&4096)&&(8,&128)&&(32,&2048)\\
&\#3&&(4,&1024)&&(64,&4096)&&(32,&2048)&&(64,&2048)&&(8&512)&&(32,&8192)\\
\hline
\end{tabular}
\end{table} 
\normalsize

\clearpage

\begin{table}[t!]
\caption{Comparison of the behavior of DFA for the pair $(4,32)$ on fractional Brownian motion simulated using Davies and Harte exact method (\texttt{dh}) and on ARFIMA$(0,d,0)$.}
\label{tab:dh_arfima0d0}
\footnotesize
\[\begin{array}{rrrrrrrrrrrrrr}
\hline
N	&		&	&	\multicolumn{11}{l}{\text{Nominal $H$}}\\
\cline{4-14}
	&		&		&	0.5	&	0.6	&	0.7	&	0.8	&	0.9	&		&	0.5	&	0.6	&	0.7	&	0.8	&	0.9\\
\cline{4-8}\cline{10-14}
	&		&		&	\multicolumn{5}{l}{\text{Bias}}	&		&	\multicolumn{5}{l}{\text{Std. dev.}}\\
\cline{1-1}\cline{3-3}\cline{4-8}\cline{10-14}	
	128	&		&	\texttt{\texttt{dh}}						&	-0.005	&	-0.013	&	-0.024	&	-0.028	&	-0.037	&		&	0.082	&	0.089	&	0.096	&	0.102	&	0.110\\
			&		&	\text{ARFIMA}(0,d,0)	&	-0.003	&	-0.030	&	-0.053	&	-0.072	&	-0.083	&		&	0.081	&	0.089	&	0.095	&	0.103	&	0.110\\
	256	&		&	\texttt{dh}						&	0.000	&	-0.009	&	-0.015	&	-0.020	&	-0.025	&		&	0.058	&	0.064	&	0.069	&	0.074	&	0.079\\
			&		&	\text{ARFIMA}(0,d,0)	&	0.000	&	-0.026	&	-0.046	&	-0.061	&	-0.074	&		&	0.058	&	0.062	&	0.067	&	0.073	&	0.077\\
	512	&		&	\texttt{dh}						&	0.002	&	-0.005	&	-0.011	&	-0.014	&	-0.018	&		&	0.041	&	0.045	&	0.049	&	0.053	&	0.056\\
			&		&	\text{ARFIMA}(0,d,0)	&	0.002	&	-0.023	&	-0.043	&	-0.059	&	-0.069	&		&	0.041	&	0.045	&	0.049	&	0.051	&	0.055\\
1024	&		&	\texttt{dh}						&	0.004	&	-0.004	&	-0.009	&	-0.012	&	-0.014	&		&	0.029	&	0.032	&	0.035	&	0.037	&	0.039\\
			&		&	\text{ARFIMA}(0,d,0)	&	0.004	&	-0.021	&	-0.042	&	-0.056	&	-0.067	&		&	0.029	&	0.031	&	0.034	&	0.036	&	0.039\\
2048	&		&	\texttt{dh}						&	0.004	&	-0.002	&	-0.007	&	-0.011	&	-0.012	&		&	0.021	&	0.023	&	0.025	&	0.026	&	0.028\\
			&		&	\text{ARFIMA}(0,d,0)	&	0.004	&	-0.021	&	-0.041	&	-0.056	&	-0.066	&		&	0.021	&	0.023	&	0.024	&	0.026	&	0.028\\
4096	&		&	\texttt{dh}						&	0.005	&	-0.003	&	-0.007	&	-0.009	&	-0.011	&		&	0.015	&	0.016	&	0.017	&	0.019	&	0.020\\
			&		&	\text{ARFIMA}(0,d,0)	&	0.004	&	-0.021	&	-0.040	&	-0.055	&	-0.065	&		&	0.015	&	0.016	&	0.017	&	0.018	&	0.019\\
8192	&		&	\texttt{dh}						&	0.004	&	-0.002	&	-0.006	&	-0.009	&	-0.011	&		&	0.010	&	0.011	&	0.012	&	0.013	&	0.014\\
			&		&	\text{ARFIMA}(0,d,0)	&	0.004	&	-0.021	&	-0.040	&	-0.055	&	-0.065	&		&	0.010	&	0.011	&	0.012	&	0.013	&	0.014\\
16384	&		&	\texttt{dh}						&	0.004	&	-0.002	&	-0.006	&	-0.009	&	-0.010	&		&	0.007	&	0.008	&	0.009	&	0.009	&	0.010\\
			&		&	\text{ARFIMA}(0,d,0)	&	0.004	&	-0.021	&	-0.040	&	-0.055	&	-0.065	&		&	0.007	&	0.008	&	0.009	&	0.009	&	0.010\\
\hline
\end{array}\]
\end{table}
\normalsize

\begin{table}[t!]
\caption{Example of the behavior of DFA for the pair $(4,32)$ on \emph{fBm} simulated by Davies and Harte exact method (\texttt{dh}) and on ARFIMA$(p,d,q)$ of length $N=8192$.}
\label{tab:dh_arfimapdq}
\small
\[\begin{array}{llrrrrrrrrrrrr}
\hline
\multicolumn{2}{l}{N=8192,(m^-,m^+)=(4,32)}&&\multicolumn{11}{l}{\text{Nominal $H$}}\\
\cline{4-14}
										&												&&0.5		&0.6		&0.7		&0.8		&0.9		&&0.5	 &0.6	 &0.7	 &0.8	 &0.9\\
\cline{4-8}\cline{10-14}
&&&\multicolumn{5}{l}{\text{Bias}}&&\multicolumn{5}{l}{\text{Std. dev.}}\\
\hline
\texttt{dh}&																&&0.004	&-0.002	&-0.006	&-0.009	&-0.011	&&0.010&0.011&0.012&0.013&0.014\\
\text{ARFIMA}(0,d,0)&												&&0,004	&-0.021	&-0.040	&-0.055	&-0.065	&&0.010&0.011&0.012&0.013&0.014\\
\text{ARFIMA}(0,d,1)&\theta=0.5							&&-0.276&-0.323	&-0.361	&-0.389	&-0.409	&&0.007&0.008&0.009&0.010&0.012\\
\text{ARFIMA}(1,d,0)&\phi=0.5								&&0.311	&0.287	&0.266	&0.246	&0.228	&&0.012&0.013&0.013&0.014&0.015\\
\text{ARFIMA}(1,d,1)&\phi=0.3,\theta=0.7		&&-0.255&-0.306	&-0.350	&-0.386	&-0.414	&&0.006&0.007&0.008&0.010&0.011\\
\text{ARFIMA}(1,d,1)&\phi=-0.3,\theta=-0.7	&&0.105	&0.078	&0.055	&0.036	&0.020	&&0.011&0.012&0.012&0.013&0.014\\
\text{ARFIMA}(1,d,1)&\phi=0.7,\theta=0.3		&&0.343	&0.328	&0.314	&0.301	&0.289	&&0.013&0.014&0.014&0.015&0.015\\
\hline
\end{array}\]
\end{table}
\normalsize

\begin{table}[t!]
\caption{Bias and standard deviation for DFA based on optimal blocks $(\tilde{m}^-,\tilde{m}^-)$ for \emph{fBm} simulated by Davies and Harte exact method (\texttt{dh}) and ARFIMA$(p,d,q)$ of length $N=8192$.}
\label{tab:dh_arfimapdq2}
\small
\[\begin{array}{llrrrrrrrrrrrr}
\hline
\multicolumn{2}{l}{N=8192}&(\tilde{m}^-,\tilde{m}^-)&\multicolumn{11}{l}{\text{Nominal $H$}}\\
\cline{4-14}
										&												&&0.5		&0.6		&0.7		&0.8		&0.9		&&0.5	 &0.6	 &0.7	 &0.8	 &0.9\\
\cline{4-8}\cline{10-14}
&&&\multicolumn{5}{l}{\text{Bias}}&&\multicolumn{5}{l}{\text{Std. dev.}}\\
\hline
\texttt{dh}&																&(4,32)			&0.004	&-0.002	&-0.006	&-0.009	&-0.011	&&0.010&0.011&0.012&0.013&0.014\\
\text{ARFIMA}(0,d,0)&												&(8,512)		&-0.009	&-0.017	&-0.021	&-0.023	&-0.024	&&0.016&0.018&0.020&0.021&0.023\\
\text{ARFIMA}(0,d,1)&\theta=0.5							&(64,2048)	&-0.058	&-0.045	&-0.036	&-0.030	&-0.027	&&0.040&0.045&0.050&0.054&0.058\\
\text{ARFIMA}(1,d,0)&\phi=0.5								&(32,1024)	&0.032	&0.023	&0.015	&0.010	&0.006	&&0.030&0.033&0.036&0.039&0.042\\
\text{ARFIMA}(1,d,1)&\phi=0.3,\theta=0.7		&(128,2048)	&-0.068	&-0.052	&-0.043	&-0.033	&-0.031	&&0.051&0.059&0.065&0.072&0.076\\
\text{ARFIMA}(1,d,1)&\phi=-0.3,\theta=-0.7	&(8,256)		&0.019	&0.007	&-0.001	&-0.007	&-0.011	&&0.015&0.017&0.018&0.020&0.021\\
\text{ARFIMA}(1,d,1)&\phi=0.7,\theta=0.3		&(32,4096)	&0.037	&0.026	&0.018	&0.011	&0.006	&&0.037&0.041&0.046&0.048&0.051\\
\hline
\end{array}\]
\end{table}
\normalsize

\clearpage

\begin{table}[t!]
\caption{Bias and standard deviation for DFA based on optimal blocks $(\tilde{m}^-,\tilde{m}^-)$ for available ARFIMA$(p,d,q)$ of length $N=2^7,\ldots,2^{13}$.}
\label{tab:dh_arfimapdq3}
\small
\[\begin{array}{lllrrrrrrrrrrrr}
\hline
&&&(\tilde{m}^-,\tilde{m}^-)&\multicolumn{11}{l}{\text{Nominal $H$}}\\
\cline{5-15}
&										&												&&0.5		&0.6		&0.7		&0.8		&0.9		&&0.5	 &0.6	 &0.7	 &0.8	 &0.9\\
\cline{5-9}\cline{11-15}
&&&&\multicolumn{5}{l}{\text{Bias}}&&\multicolumn{5}{l}{\text{Std. dev.}}\\
\hline
\multicolumn{2}{l}{N=128}\\
&\text{ARFIMA}(0,d,0)&												&(8,128)		&-0.037&-0.053&-0.068&-0.077&-0.087&&0.103&0.115&0.125&0.136&0.145\\
&\text{ARFIMA}(0,d,1)&\theta=0.5							&(8,128)		&-0.242&-0.263&-0.275&-0.276&-0.278&&0.080&0.096&0.111&0.125&0.138\\
&\text{ARFIMA}(1,d,0)&\phi=0.5								&(8,128)		&0.124&0.101&0.079&0.059&0.043&&0.114&0.123&0.131&0.142&0.150\\
&\text{ARFIMA}(1,d,1)&\phi=0.3.\theta=0.7			&(8,128)		&-0.275&-0.305&-0.326&-0.340&-0.345&&0.070&0.087&0.102&0.115&0.130\\
&\text{ARFIMA}(1,d,1)&\phi=-0.3.\theta=-0.7		&(8,128)		&0.000&-0.020&-0.039&-0.053&-0.061&&0.106&0.117&0.126&0.136&0.144\\
&\text{ARFIMA}(1,d,1)&\phi=0.7.\theta=0.3			&(8,128)		&0.199&0.177&0.156&0.138&0.114&&0.119&0.129&0.137&0.145&0.149\\
\multicolumn{2}{l}{N=256}\\
&\text{ARFIMA}(0,d,0)&												&(16,256)		&-0.035&-0.043&-0.054&-0.062&-0.071&&0.101&0.113&0.124&0.135&0.146\\
&\text{ARFIMA}(0,d,1)&\theta=0.5							&(16,256)		&-0.185&-0.188&-0.185&-0.181&-0.170&&0.086&0.103&0.117&0.130&0.141\\
&\text{ARFIMA}(1,d,0)&\phi=0.5								&(16,256)		&0.060&0.038&0.021&0.005&-0.009&&0.109&0.119&0.130&0.139&0.146\\
&\text{ARFIMA}(1,d,1)&\phi=0.3.\theta=0.7			&(16,256)		&-0.234&-0.246&-0.248&-0.246&-0.238&&0.079&0.096&0.111&0.125&0.138\\
&\text{ARFIMA}(1,d,1)&\phi=-0.3.\theta=-0.7		&(16,256)		&-0.016&-0.032&-0.041&-0.051&-0.061&&0.103&0.115&0.123&0.136&0.146\\
&\text{ARFIMA}(1,d,1)&\phi=0.7.\theta=0.3			&(16,256)		&0.121&0.099&0.080&0.061&0.045&&0.114&0.121&0.131&0.140&0.149\\
\multicolumn{2}{l}{N=512}\\
&\text{ARFIMA}(0,d,0)&												&(16,512)		&-0.028&-0.038&-0.044&-0.050&-0.056&&0.079&0.087&0.096&0.104&0.110\\
&\text{ARFIMA}(0,d,1)&\theta=0.5							&(16,512)		&-0.158&-0.154&-0.150&-0.141&-0.131&&0.071&0.083&0.092&0.102&0.109\\
&\text{ARFIMA}(1,d,0)&\phi=0.5								&(16,512)		&0.048&0.030&0.012&0.001&-0.010&&0.081&0.089&0.098&0.105&0.114\\
&\text{ARFIMA}(1,d,1)&\phi=0.3,\theta=0.7			&(16,512)		&-0.204&-0.209&-0.206&-0.199&-0.191&&0.067&0.079&0.090&0.099&0.108\\
&\text{ARFIMA}(1,d,1)&\phi=-0.3,\theta=-0.7		&(16,512)		&-0.013&-0.025&-0.036&-0.042&-0.051&&0.079&0.087&0.096&0.104&0.111\\
&\text{ARFIMA}(1,d,1)&\phi=0.7,\theta=0.3			&(16,512)		&0.099&0.082&0.063&0.048&0.031&&0.084&0.090&0.098&0.106&0.113\\
\multicolumn{2}{l}{N=1024}\\
&\text{ARFIMA}(0,d,0)&												&(16,1024)	&-0.023&-0.030&-0.036&-0.040&-0.047&&0.062&0.069&0.075&0.082&0.088\\
&\text{ARFIMA}(0,d,1)&\theta=0.5							&(16,1024)	&-0.132&-0.128&-0.121&-0.115&-0.107&&0.058&0.068&0.074&0.080&0.087\\
&\text{ARFIMA}(1,d,0)&\phi=0.5								&(16,1024)	&0.037&0.024&0.011&0.001&-0.011&&0.064&0.070&0.077&0.083&0.089\\
&\text{ARFIMA}(1,d,1)&\phi=0.3,\theta=0.7			&(16,1024)	&-0.176&-0.175&-0.170&-0.162&-0.153&&0.056&0.066&0.073&0.081&0.087\\
&\text{ARFIMA}(1,d,1)&\phi=-0.3,\theta=-0.7		&(16,1024)	&-0.012&-0.020&-0.028&-0.035&-0.041&&0.062&0.070&0.076&0.082&0.088\\
&\text{ARFIMA}(1,d,1)&\phi=0.7,\theta=0.3			&(16,1024)	&0.082&0.065&0.049&0.037&0.025&&0.064&0.071&0.077&0.083&0.087\\
\multicolumn{2}{l}{N=2048}\\
&\text{ARFIMA}(0,d,0)&												&(32,1024)	&-0.015&-0.020&-0.023&-0.025&-0.031&&0.057&0.065&0.070&0.076&0.081\\
&\text{ARFIMA}(0,d,1)&\theta=0.5							&(32,1024)	&-0.099&-0.088&-0.078&-0.070&-0.063&&0.054&0.063&0.068&0.075&0.080\\
&\text{ARFIMA}(1,d,0)&\phi=0.5								&(32,1024)	&0.025&0.012&0.003&-0.005&-0.012&&0.058&0.064&0.071&0.076&0.081\\
&\text{ARFIMA}(1,d,1)&\phi=0.3,\theta=0.7			&(32,1024)	&-0.142&-0.132&-0.120&-0.106&-0.096&&0.051&0.061&0.068&0.074&0.079\\
&\text{ARFIMA}(1,d,1)&\phi=-0.3,\theta=-0.7		&(32,1024)	&-0.009&-0.013&-0.020&-0.024&-0.027&&0.057&0.064&0.070&0.076&0.081\\
&\text{ARFIMA}(1,d,1)&\phi=0.7,\theta=0.3			&(32,1024)	&0.055&0.042&0.029&0.019&0.010&&0.059&0.065&0.072&0.076&0.082\\
\multicolumn{2}{l}{N=4096}\\
&\text{ARFIMA}(0,d,0)&												&(32,2048)	&-0.013&-0.017&-0.019&-0.022&-0.024&&0.045&0.050&0.055&0.060&0.064\\
&\text{ARFIMA}(0,d,1)&\theta=0.5							&(32,2048)	&-0.081&-0.072&-0.062&-0.055&-0.050&&0.044&0.050&0.055&0.058&0.063\\
&\text{ARFIMA}(1,d,0)&\phi=0.5								&(32,2048)	&0.020&0.010&0.002&-0.004&-0.011&&0.045&0.051&0.056&0.060&0.065\\
&\text{ARFIMA}(1,d,1)&\phi=0.3,\theta=0.7			&(32,2048)	&-0.119&-0.108&-0.097&-0.085&-0.077&&0.044&0.049&0.055&0.059&0.064\\
&\text{ARFIMA}(1,d,1)&\phi=-0.3,\theta=-0.7		&(32,2048)	&-0.007&-0.011&-0.016&-0.020&-0.023&&0.045&0.050&0.055&0.060&0.064\\
&\text{ARFIMA}(1,d,1)&\phi=0.7,\theta=0.3			&(32,2048)	&0.045&0.033&0.023&0.015&0.007&&0.047&0.051&0.056&0.059&0.064\\
\multicolumn{2}{l}{N=8192}\\
&\text{ARFIMA}(0,d,0)&												&(64,2048)	&-0.008&-0.011&-0.012&-0.014&-0.016&&0.041&0.046&0.050&0.054&0.057\\
&\text{ARFIMA}(0,d,1)&\theta=0.5							&(64,2048)	&-0.058&-0.045&-0.036&-0.030&-0.027&&0.040&0.045&0.050&0.054&0.058\\
&\text{ARFIMA}(1,d,0)&\phi=0.5								&(64,2048)	&0.012&0.005&-0.001&-0.005&-0.008&&0.042&0.046&0.051&0.054&0.058\\
&\text{ARFIMA}(1,d,1)&\phi=0.3,\theta=0.7			&(64,2048)	&-0.090&-0.073&-0.062&-0.049&-0.043&&0.039&0.045&0.050&0.054&0.058\\
&\text{ARFIMA}(1,d,1)&\phi=-0.3,\theta=-0.7		&(64,2048)	&-0.004&-0.007&-0.011&-0.012&-0.014&&0.042&0.046&0.050&0.054&0.058\\
&\text{ARFIMA}(1,d,1)&\phi=0.7,\theta=0.3			&(64,2048)	&0.029&0.019&0.012&0.005&0.001&&0.042&0.046&0.051&0.055&0.058\\
\hline
\end{array}\]
\end{table}
\normalsize

\clearpage

\begin{figure}[b!]
	\centering
		\includegraphics[width=.8\textwidth]{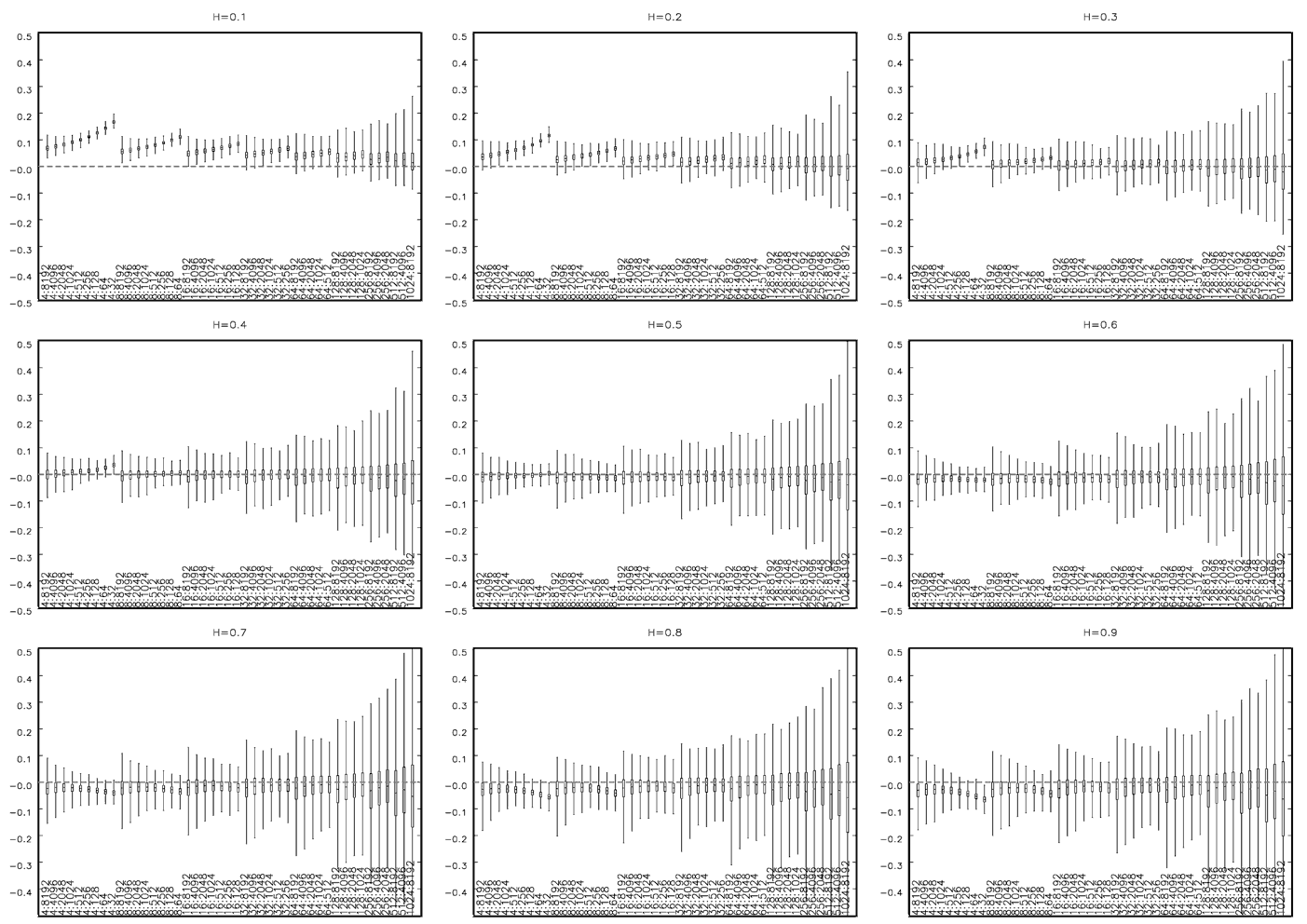}
	\caption{Boxplots for 10,000 ARFIMA(0,d,0) series of length $N=8192$. On X axis --- different block combinations, starting from 4 obs. in the shortest block and $N$ in the longest one. Longest blocks are cut first till at least four blocks are left. On Y axis --- deviation of the nominal value.}
	\label{fig:arfima0d0}
\end{figure}

\begin{figure}[b!]
	\centering
		\includegraphics[width=.85\textwidth]{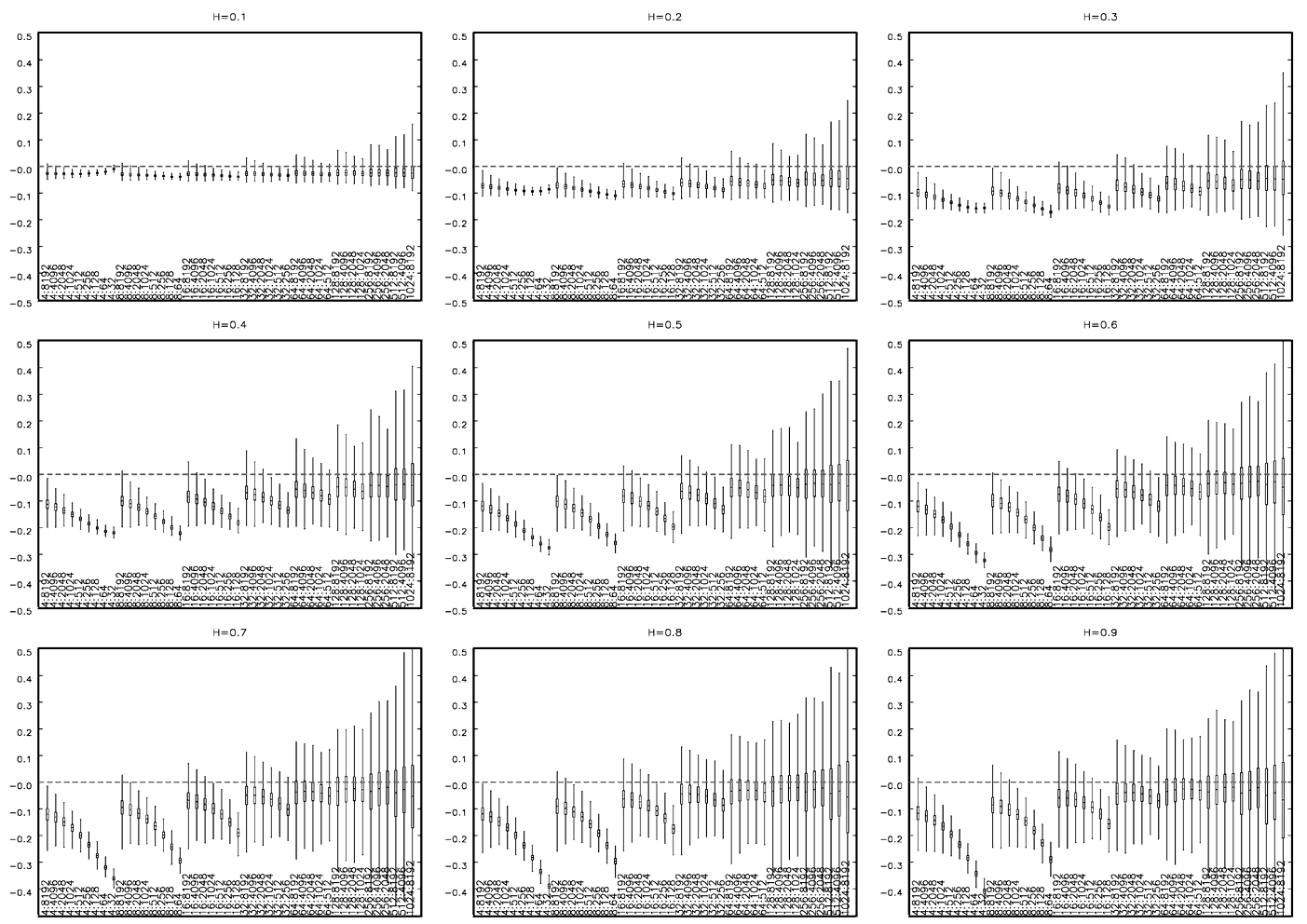}
	\caption{Same as Figure \ref{fig:arfima0d0} but ARFIMA(0,d,1), $\theta=0.5$.}
	\label{fig:arfima0d1}
\end{figure}

\clearpage

\begin{figure}[b!]
	\centering
		\includegraphics[width=.85\textwidth]{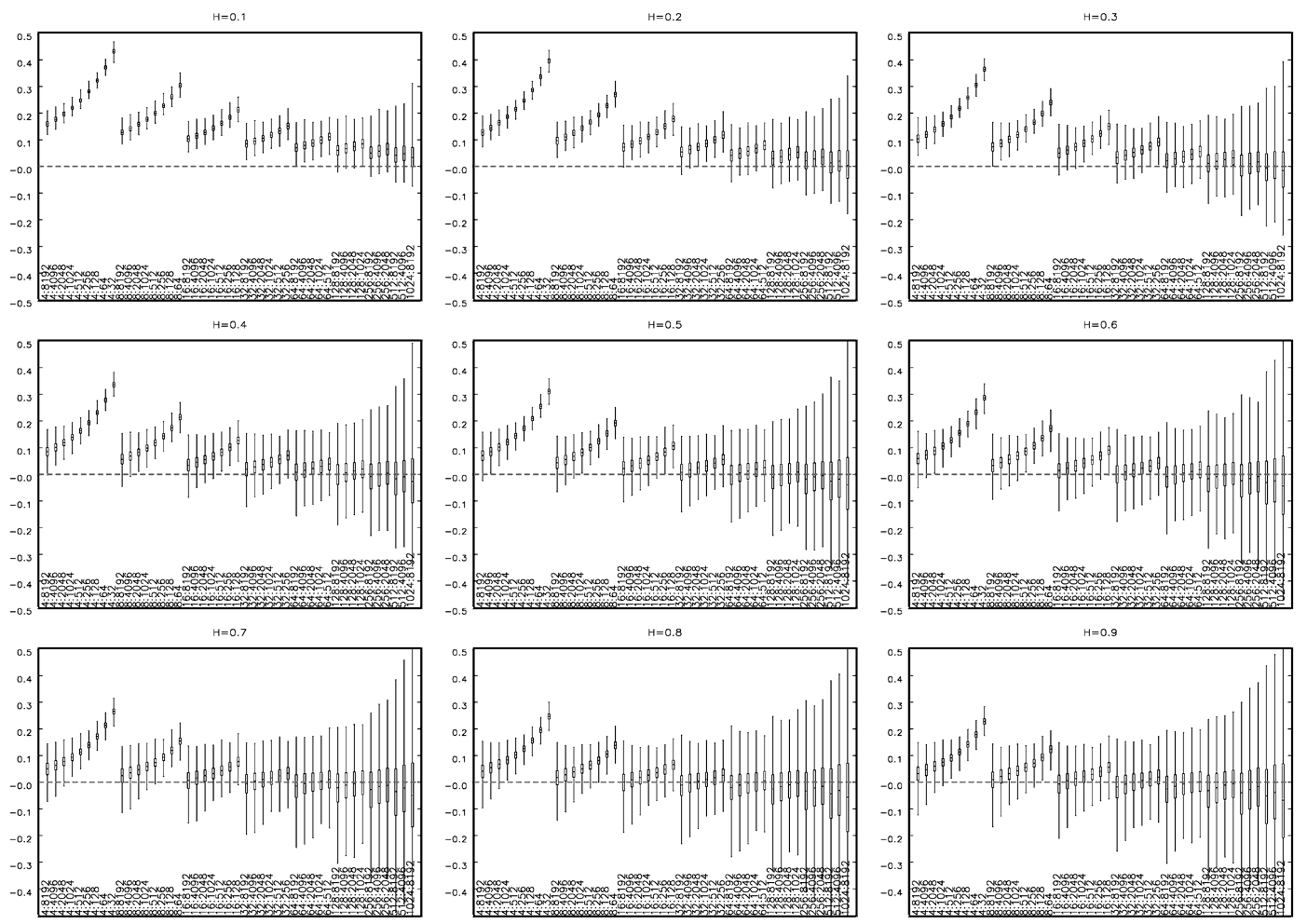}
	\caption{Same as Figure \ref{fig:arfima0d0} but ARFIMA(1,d,0), $\phi=0.5$.}
	\label{fig:arfima1d0}
\end{figure}

\begin{figure}[b!]
	\centering
		\includegraphics[width=.85\textwidth]{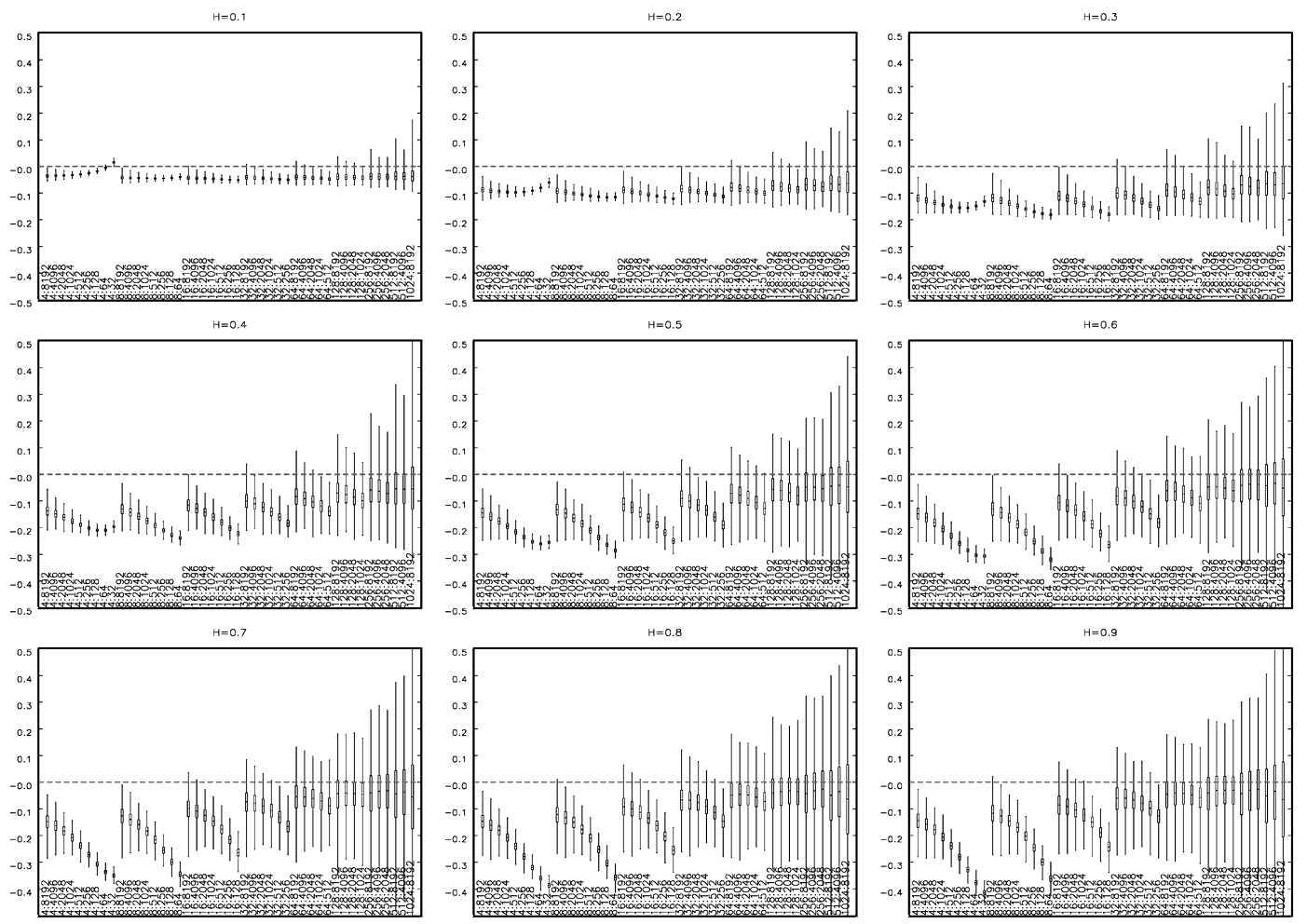}
	\caption{Same as Figure \ref{fig:arfima0d0} but ARFIMA(1,d,1), $\phi=0.3$, $\theta=0.7$.}
	\label{fig:arfima1d1a}
\end{figure}

\clearpage

\begin{figure}[b!]
	\centering
		\includegraphics[width=.85\textwidth]{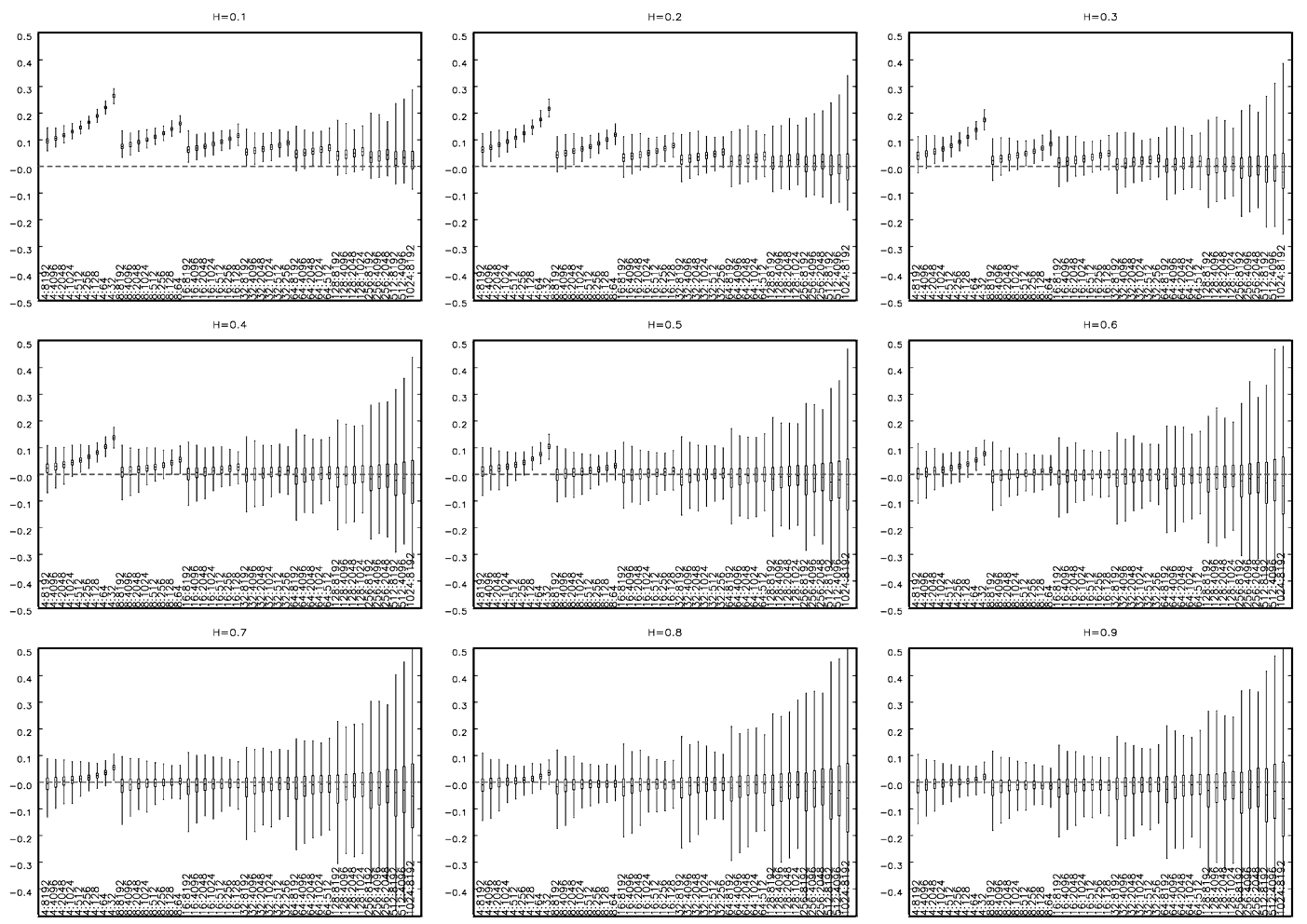}
	\caption{Same as Figure \ref{fig:arfima0d0} but ARFIMA(1,d,1), $\phi=-0.3$, $\theta=-0.7$.}
	\label{fig:arfima1d1b}
\end{figure}

\begin{figure}[b!]
	\centering
		\includegraphics[width=.85\textwidth]{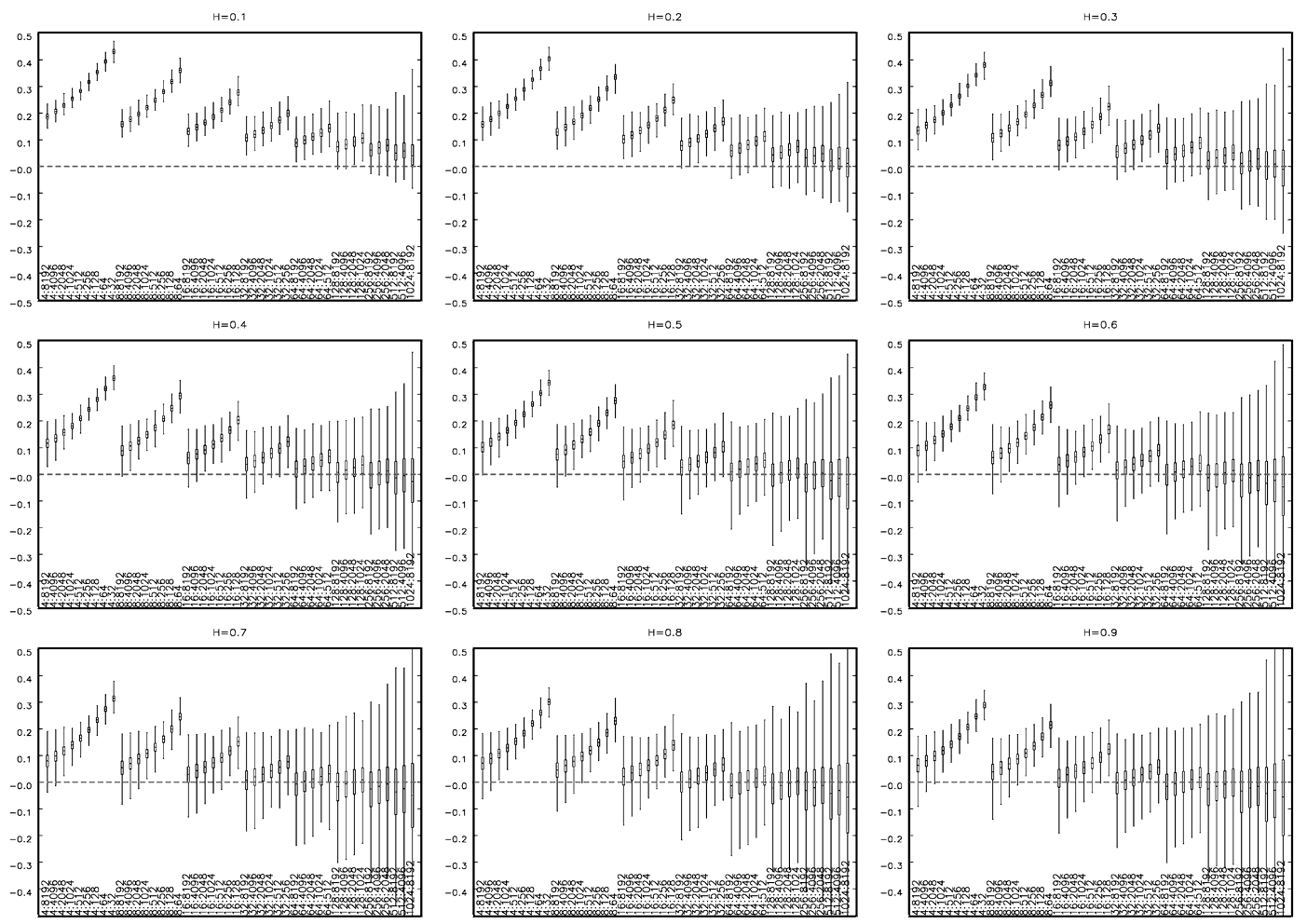}
	\caption{Same as Figure \ref{fig:arfima0d0} but ARFIMA(1,d,1), $\phi=0.7$, $\theta=0.3$.}
	\label{fig:arfima1d1c}
\end{figure}

\clearpage

%







\end{document}